\documentclass[final,1p,times]{elsarticle}

\usepackage{graphicx}

\usepackage{amsmath}
\usepackage{amssymb}

\usepackage[autolanguage]{numprint}

\usepackage{natbib}

\biboptions{sort&compress}

\usepackage[colorlinks]{hyperref}

\journal{Journal of Computational Physics}

\begin{document}

\begin{frontmatter}



\title{A fast lattice Green's function method for solving viscous incompressible flows on unbounded domains}


\author{Sebastian Liska\corref{cor1}}
\ead{sliska@caltech.edu}
\author{Tim Colonius}
\ead{colonius@caltech.edu}
\cortext[cor1]{Corresponding author}
\address{Division of Engineering and Applied Science, California Institute of Technology, Pasadena, CA 91125, USA}

\begin{abstract}
A computationally efficient method for solving three-dimensional, viscous, incompressible flows on unbounded domains is presented.
The method formally discretizes the incompressible Navier-Stokes equations on an unbounded staggered Cartesian grid.
Operations are limited to a finite computational domain through a lattice Green's function technique.
This technique obtains solutions to inhomogeneous \emph{difference} equations through the discrete convolution of source terms with the fundamental solutions of the \emph{discrete} operators.
The differential algebraic equations describing the temporal evolution of the discrete momentum equation and incompressibility constraint are numerically solved by combining an integrating factor technique for the viscous term and a half-explicit Runge-Kutta scheme for the convective term.
A projection method that exploits the mimetic and commutativity properties of the discrete operators is used to efficiently solve the system of equations that arises in each stage of the time integration scheme.
Linear complexity, fast computation rates, and parallel scalability are achieved using recently developed fast multipole methods for difference equations.
The accuracy and physical fidelity of solutions is verified through numerical simulations of vortex rings.
\end{abstract}


\end{frontmatter}

\section{Introduction}
\label{sec:intro}
Numerical simulations of viscous, incompressible flows on unbounded fluid domains require numerical techniques that can accurately approximate unbounded computational domains using only a finite number of operations.
Spatial truncation and artificial boundary conditions have been developed for this purpose but they can adversely affect the accuracy of the solution and even change the dynamics of the flow \cite{tsynkov1998,colonius2004,pradeep2004,dong2014}.
Furthermore, minimizing the error due to artificial boundaries by employing large computational domains increases the number of computational elements and often requires the use of solvers that are less efficient than those used on regular grids (e.g. FFT techniques, multigrid, etc.).

Recently, fast multipole methods (FMMs) for solving constant coefficient elliptic \emph{difference} equations on unbounded regular grids have been developed for 2D \cite{gillman2010,gillman2014} and 3D \cite{liska2014} problems.
These methods obtain solutions to inhomogeneous \emph{difference} equations by using fast summation techniques to evaluate the discrete convolution of source terms with the fundamental solutions of the \emph{discrete} operators.
The fundamental solutions of discrete operators on unbounded regular grids, or lattices, are also referred to as lattice Green's functions (LGFs).

Similar to particle and vortex methods, e.g.  \cite{leonard1980,greengard1987,winckelmans1993,warren1993,cheng1999,ploumhans2000,cottet2000,ying2004,winckelmans2004,cocle2008,chatelain2010,rasmussen2011,hejlesen2013} and references therein, the LGF techniques discussed in \cite{gillman2010,gillman2014,liska2014} have efficient nodal distributions and automatically enforce free-space boundary conditions.
As a result, needlessly large computational domains and artificial boundary conditions can be avoided when solving flows on unbounded regular grids by using LGF techniques to compute the action of solution operators.
A significant advantage of recently developed particle and vortex methods is their ability to efficiently solve large scale problems relevant to 3D incompressible flows using fast, parallel methods based on techniques such as tree-codes, FMMs, dynamic error estimators, hybrid Eulerian-Lagrangian formulations, hierarchical grids, FFT methods, and domain decomposition techniques \cite{warren1993,cheng1999,ploumhans2000,ying2004,cocle2008,chatelain2010,rasmussen2011,hejlesen2013}.
It is demonstrated in \cite{liska2014} that LGF FMMs can achieve computational rates and parallel scaling for 3D discrete (7-pt Laplacian) Poisson problems comparable to existing fast 3D Poisson solvers.

The present formulation numerically solves the incompressible Navier-Stokes equations expressed in the non-dimensional form given by
\begin{subequations}
  \begin{align}
    \frac{ \partial \mathbf{u} }{ \partial t }
      + \mathbf{u} \cdot \nabla \mathbf{u}
      &= - \nabla p + \frac{1}{\text{Re}} \nabla^{2} \mathbf{u}, \label{eq:ns-mom} \\
    \nabla \cdot \mathbf{u} &=0,\label{eq:ns-cont}
  \end{align}
  \label{eq:ns}%
\end{subequations}
where $\mathbf{u}$, $p$, and $\text{Re}$ correspond to the velocity, the pressure, and the Reynolds number, respectively.
The equations are defined on an unbounded domain in all directions, and are subject to the boundary conditions
\begin{equation}
  \mathbf{u} \left( \mathbf{x}, t \right) \rightarrow \mathbf{u}_\infty\left( t \right) \,\, \text{as} \,\, \left|\mathbf{x}\right| \rightarrow \infty,
  \label{eq:ns_bcs}
\end{equation}
where $\mathbf{u}_\infty$ is a known time-dependent function.
We limit our attention to flows in which the vorticity, $\boldsymbol{\omega} = \nabla \times \mathbf{u}$, decay exponentially fast as $\left|\mathbf{x}\right| \rightarrow \infty$.

The present formulation is simplified by considering the evolution of the velocity perturbation, $\mathbf{u}^\prime\left( \mathbf{x}, t \right) = \mathbf{u}\left( \mathbf{x}, t \right) - \mathbf{u}_\infty\left( t \right)$, and pressure perturbation, $p^\prime\left( \mathbf{x}, t \right) = p\left( \mathbf{x}, t \right) - p_\infty\left( \mathbf{x}, t \right)$.
The freestream pressure, $p_\infty$, is given by
\begin{equation}
  p_\infty\left(\mathbf{x},t\right) = \frac{d \mathbf{u}_\infty } { dt } \cdot \mathbf{x},
  \label{eq:press_infty}
\end{equation}
where we have taken the arbitrary time-dependent constant to be zero.
Subtracting the uniform freestream equations from Eq.~\eqref{eq:ns} yields
\begin{equation}
  \begin{split}
    \frac{ \partial \mathbf{u}^\prime }{ \partial t } \
      + \left( \mathbf{u}^\prime + \mathbf{u}_\infty \right) \cdot \nabla \mathbf{u}^\prime \
      = - \nabla p^\prime + \frac{1}{\text{Re}} \nabla^{2} \mathbf{u}^\prime,
  \end{split} \quad
  \begin{split}
    \nabla \cdot \mathbf{u}^\prime = 0,
  \end{split}
  \label{eq:pns}
\end{equation}
subject to the boundary conditions $\mathbf{u}^\prime \left( \mathbf{x}, t \right) \rightarrow 0$ as $\left|\mathbf{x}\right| \rightarrow \infty$.
The boundary conditions on $\mathbf{u}^\prime$ and the irrotational nature of the flow at large distances imply that $p^\prime$ is subject to the compatibility condition%
\footnote{%
In the absence of sources and sinks, the velocity of an irrotational flow subject to zero boundary conditions at infinity is given by $\mathbf{v} = \nabla \phi$, where the leading order term of $\phi$ is $-\mathbf{M}\cdot\mathbf{x}/r^3$ \cite{saffman1992}. Consequently, $p = -\left( \frac{ \partial \phi }{ \partial t } + \frac{1}{2} | \nabla \phi |^2 \right) \rightarrow 0$ as $r \rightarrow \infty$, where we have taken the arbitrary time-dependent constant to be zero.
}
\begin{equation}
p^\prime \left( \mathbf{x}, t \right) \rightarrow 0\,\,
\text{as}\,\, \left|\mathbf{x}\right| \rightarrow \infty.
\label{eq:press_compat}
\end{equation}

The remainder of the paper is organized as follows.
In Section~\ref{sec:spatial}, we describe the spatial discretization of the governing equations on formally unbounded staggered Cartesian grids and discuss LGF techniques that can be used to obtain fast solutions to the associated discrete elliptic problems.
Additionally, we present an integrating factor technique that facilitates the implementation of efficient, robust time integration schemes.
In Section~\ref{sec:temporal}, the system of differential algebraic equations (DAEs) resulting from the spatial discretization and integrating factor techniques is numerically solved using a half-explicit Runge-Kutta method.
We show that the linear systems of equations that arise at each stage of the time integration scheme can be efficiently solved, without splitting errors or additional stability constraints, by a fast projection method based on LGF techniques and the properties of the discrete operators.
In Section~\ref{sec:truncation}, we demonstrate that an adaptive block-structured grid padded with appropriately sized buffer regions can be used to efficiently compute numerical solutions to a prescribed tolerance.
In Section~\ref{sec:algorithm}, we summarize the algorithm and discuss a few practical considerations including computational costs and performance optimization.
Finally, in Section~\ref{sec:verif}, we perform numerical experiments on vortex rings to verify the present formulation.

\section{Spatial discretization}
\label{sec:spatial}

\subsection{Unbounded staggered Cartesian grids}
\label{sec:spatial_discrete}

\begin{figure}[htbp]
  \begin{center}
    \includegraphics[width=0.75\textwidth]{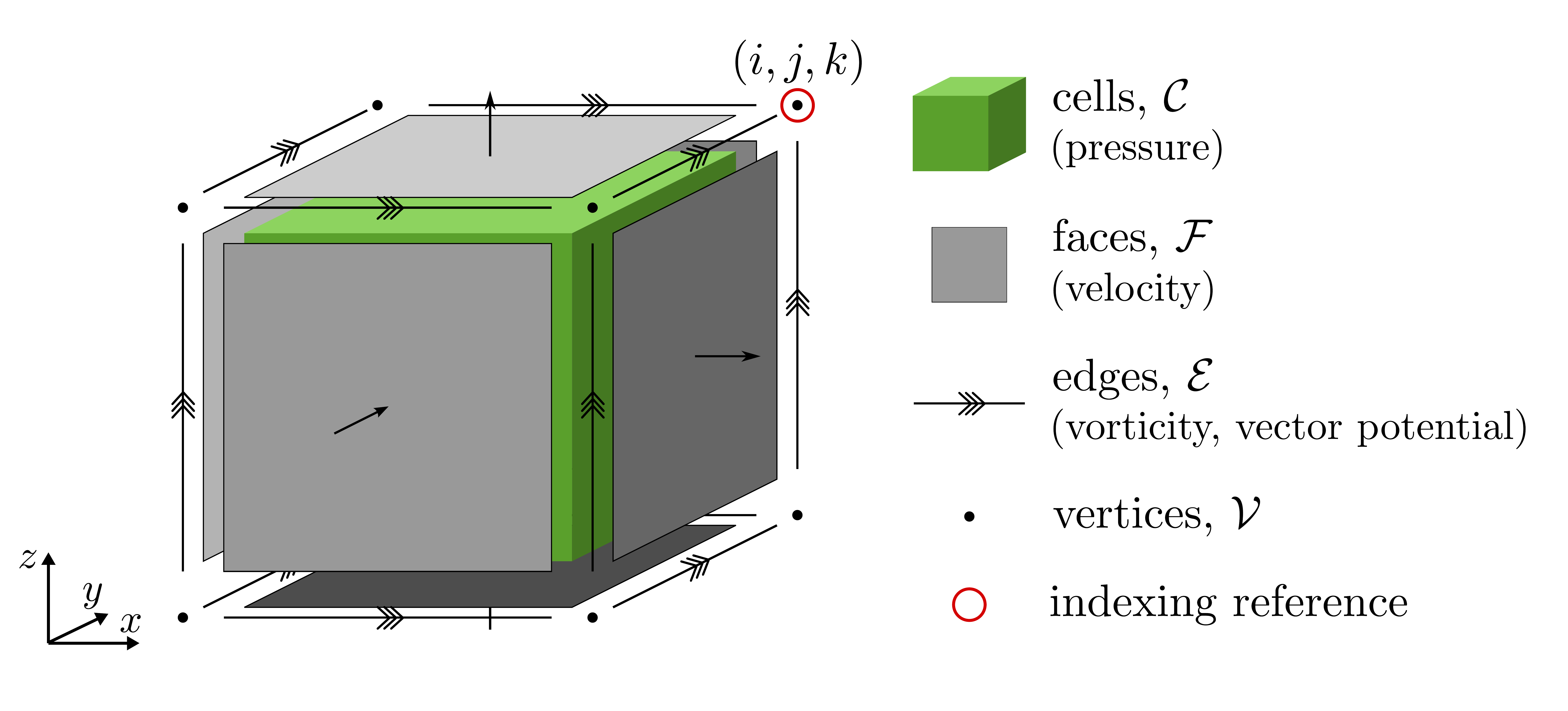}
  \end{center}
  \caption{%
  Unit cell of the staggered Cartesian grid.
  The vertex enclosed by the circle corresponds to the $(i,j,k)$ vertex.
  The $(i,j,k)$ cell, faces, and edges correspond to the depicted elements intersecting the $(i,j,k)$ vertex.
  There are three faces and edges per vertex.
  The superscript ``$(q)$'' is used to denote faces (edges) normal (parallel) to $x_q$ axis.
  \label{fig:grid-cell}
  }
\end{figure}

In this section we describe the discretization of Eq.~\eqref{eq:pns} on a formally unbounded staggered Cartesian grid.
Figure~\ref{fig:grid-cell} depicts our staggered grid, which consists of cells ($\mathcal{C}$) and vertices ($\mathcal{V}$) that house scalar quantities, and faces ($\mathcal{F}$) and edges ($\mathcal{E}$) that house vector quantities.
The notation $\mathbb{R}^{\mathcal{Q}}$ denotes the set of real-valued grid functions with values defined on $\mathcal{Q}\in\{\mathcal{C},\mathcal{F},\mathcal{E},\mathcal{V}\}$.
The value of a grid function $\mathsf{q}$ evaluated at $\mathbf{n}=(i,j,k)\in\mathbb{Z}^3$ is given by $\mathsf{q}(\mathbf{n})$ and $\mathsf{q}_{i,j,k}$.
For the case of a vector-valued grid function $\mathsf{q}$, i.e. $\mathsf{q}\in\mathbb{R}^{\mathcal{F}}$ or $\mathsf{q}\in\mathbb{R}^{\mathcal{E}}$, $\mathsf{q}^{(k)}(\mathbf{n})$ denotes the component of $\mathsf{q}(\mathbf{n})$ in the $k$-th direction.

The spatial discretization of Eq.~\eqref{eq:pns} is performed using the techniques of \citet{nicolaides1997}, and \citet{zhang2002}.
The resulting discrete operators are similar or equivalent to those obtained from standard second-order finite-volume or finite-difference schemes, e.g. \cite{harlow1965}.
Yet we refer to the more general techniques of \cite{nicolaides1997} and \cite{zhang2002} since their discussions emphasize many of the algebraic properties of the discrete operators used by the present formulation.
For convenience, point-operator representations of the discrete operators are included in \ref{app:oprs}.

The semi-discrete system of equations obtained from the spatial discretization of Eq.~\eqref{eq:pns} is
\begin{equation}
  \begin{split}
    \frac{ d \mathsf{u} }{ d t } \
      + \mathsf{N}( \mathsf{u} + \mathsf{u}_\infty ) \
      = -\mathsf{G} \mathsf{p} + \frac{1}{\text{Re}} \mathsf{L}_{\mathcal{F}} \mathsf{u},
    \end{split} \quad
    \begin{split}
      \overline{\mathsf{D}} \mathsf{u} = 0,
  \end{split}
  \label{eq:dns}
\end{equation}
where $\mathsf{u}\in\mathbb{R}^\mathcal{F}\times\mathbb{R}$ and $\mathsf{p}\in\mathbb{R}^\mathcal{C}\times\mathbb{R}$ denote the time-dependent grid functions associated with the discrete velocity and pressure perturbation fields, respectively.%
\footnote{%
In order to avoid a cumbersome notation, the prime symbols, ${}^\prime$, are omitted from variables denoting grid functions associated with the perturbations of the discrete velocity and pressure fields.
}
The time-dependent grid function $\mathsf{u}_\infty\in\mathbb{R}^\mathcal{F}\times\mathbb{R}$ is constant in space with values given by $\mathsf{u}_\infty(\mathbf{n},t) = \mathbf{u}_\infty(t)$.
Discrete operators $\mathsf{G} : \mathbb{R}^\mathcal{C} \mapsto \mathbb{R}^\mathcal{F}$, $\overline{\mathsf{D}} : \mathbb{R}^\mathcal{F} \mapsto \mathbb{R}^\mathcal{C}$, and $\mathsf{L}_{\mathcal{F}} : \mathbb{R}^\mathcal{F} \mapsto \mathbb{R}^\mathcal{F}$ correspond to the discretizations of the gradient, divergence, and vector Laplacian operators, respectively.
Finally, $\mathsf{N}:\mathbb{R}^\mathcal{F} \mapsto \mathbb{R}^\mathcal{F}$ denotes the discrete nonlinear operator approximating the convective term, i.e. $\mathsf{N}(\mathsf{u}+\mathsf{u}_\infty) \approx \left( \mathbf{u}^\prime + \mathbf{u}_\infty \right) \cdot \nabla \left( \mathbf{u}^\prime + \mathbf{u}_\infty \right) = \left( \mathbf{u}^\prime + \mathbf{u}_\infty \right) \cdot \nabla \mathbf{u}^\prime$.%
\footnote{%
No particular form (e.g. convection, rotational, divergence, skew-symmetric) or discretization scheme for the convection term is assumed by Eq.~\eqref{eq:dns}.
}

In addition to the aforementioned discrete operators, the subsequent discussion makes use of the discrete gradient operator $\overline{\mathsf{G}}:\mathbb{R}^\mathcal{V} \mapsto \mathbb{R}^\mathcal{E}$, the discrete curl operators $\mathsf{C}:\mathbb{R}^\mathcal{F} \mapsto \mathbb{R}^\mathcal{E}$ and $\overline{\mathsf{C}}:\mathbb{R}^\mathcal{E} \mapsto \mathbb{R}^\mathcal{F}$, and the discrete Laplacian operators $\mathsf{L}_\mathcal{Q}:\mathbb{R}^\mathcal{Q} \mapsto \mathbb{R}^\mathcal{Q}$, where $\mathcal{Q}\in\{\mathcal{C},\mathcal{E},\mathcal{V}\}$.
A summary of all the discrete vector operators and their definitions is also provided in \ref{app:oprs}.

The choice of discretization technique yields a numerical scheme with the following properties:
\begin{list}{\textbullet}{\leftmargin=1em \itemindent=0em}
  \item \emph{Second-order accuracy}: all discrete operators are second-order accurate in space.
  \item \emph{Conservation properties}: using appropriate discretizations of the nonlinear convective term leads to a scheme that conserves momentum, kinetic energy, and circulation in the absence of time-differencing errors and viscosity \cite{lilly1965,morinishi1998,zhang2002}.
  The benefits of discrete conservation properties related to numerical stability and physical fidelity are discussed in the review by \citet{perot2011} and references therein.
  \item \emph{Mimetic properties}: discrete operators and their corresponding vector calculus operators satisfy similar symmetry and orthogonality properties in addition to similar integration by parts formulas \cite{nicolaides1991,nicolaides1997,perot2000,zhang2002}.
  Specific properties pertinent to the discussion of the present method are:
  \begin{subequations}
    \begin{gather}
      \overline{\mathsf{D}} = -\mathsf{G}^\dagger, \quad
      \overline{\mathsf{C}} = \mathsf{C}^\dagger, \quad
      \overline{\mathsf{G}} = -\mathsf{D}^\dagger,
    \\
      \text{Im}( \mathsf{G} ) = \text{Null}( \mathsf{C} ), \quad
      \text{Im}( \mathsf{C} ) = \text{Null}( \mathsf{D} ),
    \\
      \mathsf{L}_\mathcal{C} = - \mathsf{G}^\dagger \mathsf{G}, \quad
      \mathsf{L}_\mathcal{F} = - \mathsf{G} \mathsf{G}^\dagger - \mathsf{C}^\dagger \mathsf{C}, \quad
      \mathsf{L}_\mathcal{E} = - \mathsf{D}^\dagger \mathsf{D} - \mathsf{C} \mathsf{C}^\dagger, \quad
      \mathsf{L}_\mathcal{V} = - \mathsf{D} \mathsf{D}^\dagger.
    \end{gather}
  \end{subequations}
 Many of the mimetic properties of discrete operators are closely related to the conservation properties \cite{nicolaides1997,zhang2002}.
  \item \emph{Commutativity properties}: on unbounded staggered grids, discrete Laplacians and integrating factors (to be introduced in Section~\ref{sec:spatial_intfact}) are able to commute with other operators in the sense $\mathsf{A} \mathsf{T}_\mathcal{X} = \mathsf{T}_\mathcal{Y} \mathsf{A}$, where $\mathsf{A}:\mathbb{R}^\mathcal{X}\mapsto\mathbb{R}^\mathcal{Y}$ is any of the previously mentioned linear operators, and $\mathsf{T}_\mathcal{X}$ ($\mathsf{T}_\mathcal{Y}$) is either the discrete Laplacian or integrating factor mapping $\mathbb{R}^\mathcal{X}$ to $\mathbb{R}^\mathcal{X}$ ($\mathbb{R}^\mathcal{Y}$ to $\mathbb{R}^\mathcal{Y}$).
  Similar commutativity properties result in discretizations of periodic domains using uniform staggered grids.
\end{list}
In subsequent sections we discuss how the mimetic and commutativity properties facilitate the construction of fast, stable methods for numerically solving Eq.~\eqref{eq:dns}.

It is convenient to define
\begin{equation}
  \mathsf{d} = \mathsf{p} \
    + \frac{1}{2} \mathsf{P}\left( \mathsf{u} + \mathsf{u}_\infty, \mathsf{u} + \mathsf{u}_\infty \right),
  \label{eq:tot_pres}
\end{equation}
where $\mathsf{P} : \mathbb{R}^\mathcal{F} \times \mathbb{R}^\mathcal{F} \mapsto \mathbb{R}^\mathcal{C}$ is an arbitrary discrete approximation of the vector dot-product, i.e. $\mathsf{P}( \mathsf{u}, \mathsf{v} ) \approx \mathbf{u} \cdot \mathbf{v}$.
The time-dependent grid function $\mathsf{d} \in \mathbb{R}^\mathcal{C} \times \mathbb{R}$ can be regarded as a discrete approximation of the total pressure perturbation, i.e. $\mathsf{d} \approx p^\prime + \frac{1}{2} | \mathbf{u}^\prime + \mathbf{u}_\infty |^2$.
Using Eq.~\eqref{eq:tot_pres}, we express Eq.~\eqref{eq:dns} as
\begin{equation}
  \begin{split}
    \frac{ d \mathsf{u} }{ d t } \
        + \tilde{\mathsf{N}}( \mathsf{u} + \mathsf{u}_\infty ) \
        = -\mathsf{G} \mathsf{d} + \frac{1}{\text{Re}} \mathsf{L}_{\mathcal{F}} \mathsf{u},
    \end{split} \quad
    \begin{split}
      \mathsf{G}^\dagger \mathsf{u} = 0,
  \end{split}
  \label{eq:dnstp}
\end{equation}
where $\tilde{\mathsf{N}}(\mathsf{v}) = \mathsf{N}(\mathsf{v}) - \frac{1}{2}\mathsf{G}\mathsf{P}(\mathsf{v},\mathsf{v})$.
Consequently, $\tilde{\mathsf{N}}( \mathsf{u} + \mathsf{u}_\infty )$ is a discrete approximation of $\boldsymbol{\omega}\times\left(\mathbf{u}+\mathbf{u}_\infty\right)$.%
\footnote{%
The discretization of Eq.~\eqref{eq:pns} naturally assumes the form given by Eq.~\eqref{eq:dnstp} if the convection term is discretized in its rotational form, $\left(\nabla \times \mathbf{v} \right) \times \mathbf{v}+\frac{1}{2}\nabla \mathbf{v}^2$, with the gradient term approximated by $\frac{1}{2}\mathsf{G}\mathsf{P}(\mathsf{v},\mathsf{v})$.
}
As will be demonstrated in Section~\ref{sec:truncation}, an advantage of using $\tilde{\mathsf{N}}(\mathsf{\mathsf{u} + \mathsf{u}_\infty})$ instead of $\mathsf{N}(\mathsf{\mathsf{u} + \mathsf{u}_\infty})$ is that the former typically has a smaller support than that of the latter, which in turn reduces the number of operations and storage required to numerically solve the flow.
We emphasize that Eq.~\eqref{eq:dnstp} is equivalent to Eq.~\eqref{eq:dns}, and no additional discretization errors have been introduced.

\subsection{Lattice Green's function techniques}
\label{sec:spatial_lgfs}

The procedure for solving difference equations on unbounded regular grids using LGFs is analogous to the procedure for solving inhomogeneous PDEs on unbounded domains using the fundamental solution of continuum operators.
As a representative example, we consider the (continuum) scalar Poisson equation
\begin{equation}
  [ \Delta u ] (\mathbf{x})
    = f (\mathbf{x}), \quad supp(f) \subseteq \Omega,
  \label{eq:poisson}
\end{equation}
where $\mathbf{x}\in\mathbb{R}$ and $\Omega$ is a bounded domain in $\mathbb{R}^3$.
The solution to Eq.~\eqref{eq:poisson} is given by
\begin{equation}
  u(\mathbf{x})
    = [ G * f ] (\mathbf{x})
    = \int_\Omega G( \mathbf{x} - \mathbf{y} ) f(\mathbf{y})\, d\mathbf{y},
\end{equation}
where $G(\mathbf{x})= -1/(4 \pi |\mathbf{x}|)$ is the fundamental solution of the Laplace operator.
Similarly, we consider the discrete scalar Poisson equation
\begin{equation}
  [ \mathsf{L}_{\mathcal{Q}} \mathbf{u} ] (\mathbf{n})
    = \mathsf{f}(\mathbf{n}), \quad supp(\mathsf{f}) \subseteq D,
  \label{eq:dpoisson}
\end{equation}
where $\mathsf{u},\mathsf{f}\in\mathbb{R}^\mathcal{Q}$, $D$ is a bounded region in $\mathbb{Z}^3$, and $\mathcal{Q}\in\{\mathcal{C},\mathcal{V}\}$.
The solution to Eq.~\eqref{eq:dpoisson} is given by
\begin{equation}
  \mathsf{u}(\mathbf{n})
    = [ \mathsf{G}_{\mathsf{L}} * \mathsf{f} ] (\mathbf{n})
    = \sum_{\mathbf{m}\in D} \mathsf{G}_{\mathsf{L}}(\mathbf{n}-\mathbf{m})
      \mathsf{f}(\mathbf{m})
  \label{eq:dpoisson_conv}
\end{equation}
where $\mathsf{G}_{\mathsf{L}}:\mathbb{Z}^3\mapsto\mathbb{R}$ is the fundamental solution, or LGF, of the discrete scalar Laplacian \cite{gillman2014,liska2014}.
Subsequently, we refer to the grid functions $\mathsf{f}$ and $\mathsf{u}$ as the source field and the induced field, respectively.

It is evident from the definitions of $\mathsf{L}_\mathcal{F}$ and $\mathsf{L}_\mathcal{E}$ that each component of a discrete vector Poisson problem corresponds to a discrete scalar Poisson problem.
As a result, the $q$-th component of solutions to Eq.~\eqref{eq:dpoisson} for $\mathcal{Q}\in\{\mathcal{F},\mathcal{E}\}$ are given by Eq.~\eqref{eq:dpoisson_conv} with $\mathsf{u} \rightarrow \mathsf{u}^{(q)}$ and $\mathsf{f} \rightarrow \mathsf{f}^{(q)}$.
Procedures for obtaining expressions for $\mathsf{G}_{\mathsf{L}}(\mathbf{n})$ are discussed in \cite{mccrea1940,duffin1958,buneman1971,martinsson2002}.
For convenience, expressions for $\mathsf{G}_{\mathsf{L}}(\mathbf{n})$ are provided in \ref{app:lgfs}.

Fast numerical methods for evaluating discrete convolutions involving LGFs have recently been proposed in 2D by \citet{gillman2014} and in 3D by \citet{liska2014}.
Here, the 3D lattice Green's function fast multipole method (LGF-FMM) of \cite{liska2014} is used to evaluate discrete convolutions involving $\mathsf{G}_{\mathsf{L}}$.
The LGF-FMM is a kernel-independent interpolation-based FMM specifically designed for solving difference equations on unbounded Cartesian grids.
In addition to its asymptotic linear algorithmic complexity, it has been shown that the LGF-FMM achieves high computation rates and good parallel scaling for the case of $\mathsf{G}_{\mathsf{L}}$ \cite{liska2014}.

As final remark, the LGF-FMM is a direct solver that computes solutions to a prescribed tolerance $\epsilon$, $\| \mathsf{y}_\text{true} - \mathsf{y} \|_\infty / \|\mathsf{y}_\text{true}\|_\infty \le \epsilon$, where $\mathsf{y}$ is the numerical solution and $\mathsf{y}_\text{true}$ is the exact solution to the system of \emph{difference} equations.
In order to obtain accurate error bounds for the LGF-FMM it is necessary to profile the method once for each kernel and scheme used.
Error estimates for the discrete 7-pt Laplace kernel and different schemes are provided in \cite{liska2014}.
In the present formulation, all instances of $\mathsf{E}_{\mathcal{Q}}$ and $\mathsf{L}_{\mathcal{Q}}^{-1}$ are computed using values of $\epsilon$ that are less than or equal to prescribed value of $\epsilon_\text{FMM}$.

\subsection{Integrating factor techniques}
\label{sec:spatial_intfact}

In this section we describe an integrating factor technique for integrating the stiff viscous term of Eq.~\eqref{eq:dnstp} analytically.
Analytical integration has the advantage of neither introducing discretization errors nor imposing stability constraints on the time marching scheme.
Integrating factor techniques for the viscous term are widely used in Fourier pseudo-spectral methods.
These methods typically compute the action of the integrating factor in Fourier-space.
In contrast, the present method computes the action of the integrating factor in real-space, since the Fourier series of an arbitrary grid function on an unbounded domain is not computationally practical.

We consider integrating factors defined as the solution operators of the discrete diffusion equation of the form
\begin{equation}
    \frac{d\mathsf{h}}{dt} = \kappa \mathsf{L}_{\mathcal{Q}} \mathsf{h}, \quad
    \mathsf{h}(\mathbf{n},t) \rightarrow \mathsf{h}_\infty(t)
    \,\, \text{as} \,\, |\mathbf{n}| \rightarrow \infty,
  \label{eq:ddiffusion}
\end{equation}
where $\kappa\in\mathbb{R}_{\ge 0}$ and $\mathsf{h}\in\mathbb{R}^\mathcal{Q}$.
As discussed in \ref{app:oprs}, the discrete Laplace operator $\mathsf{L}_\mathcal{Q}$ is diagonalized by the Fourier series operator $\mathfrak{F}_\mathcal{Q}$,
\begin{equation}
  (\Delta x)^2 \mathsf{L}_\mathcal{Q} = \mathfrak{F}^{-1}_\mathcal{Q}
    \sigma^{\mathsf{L}}_\mathcal{Q} \mathfrak{F}_\mathcal{Q},
\end{equation}
where $\sigma^{\mathsf{L}}_\mathcal{Q}(\boldsymbol{\xi})$ for $\boldsymbol{\xi}\in(\pi,\pi)^3$ is the spectrum of $(\Delta x)^2 \mathsf{L}_\mathcal{Q}$.
Next, we define the exponential of the $\mathsf{L}_\mathcal{Q}$ as
\begin{equation}
  \mathsf{E}_{\mathcal{Q}}( \alpha ) = \mathfrak{F}^{-1}_\mathcal{Q}
    \exp(\alpha \sigma^{\mathsf{L}}_\mathcal{Q} ) \mathfrak{F}_\mathcal{Q},
  \label{eq:if_def}
\end{equation}
where $\alpha=\kappa (t-\tau) / (\Delta x)^2$.
An immediate consequence of Eq.~\eqref{eq:if_def} is that
\begin{equation}
  \frac{d}{d\alpha} \mathsf{E}_{\mathcal{Q}}( \alpha )
  = \mathfrak{F}^{-1}_\mathcal{Q} \sigma^{\mathsf{L}}_\mathcal{Q}
    \exp(\alpha \sigma^{\mathsf{L}}_\mathcal{Q}) \mathfrak{F}^{-1}_\mathcal{Q}
  = \mathsf{L}_{\mathcal{Q}} \mathsf{E}_{\mathcal{Q}}(\alpha)
  = \mathsf{E}_{\mathcal{Q}}(\alpha) \mathsf{L}_{\mathcal{Q}},
  \label{eq:if_dt}
\end{equation}
which implies that the solution to Eq.~\eqref{eq:ddiffusion} is given by
\begin{equation}
  \mathsf{h}(\mathbf{n},t) = \left[
    \mathsf{E}_{\mathcal{Q}} \left(
      \frac{\kappa (t-\tau) }{ (\Delta x)^2 } \right) \mathsf{h}_\tau \right]
      (\mathbf{n},t),
  \quad t \ge \tau, \quad \forall \mathbf{n} \in \mathbb{Z}^{3},
  \label{eq:ddifussion_soln}
\end{equation}
where $\mathsf{h}(\mathbf{n},\tau) = \mathsf{h}_\tau(\mathbf{n})$.

We now consider using $\mathsf{E}_{\mathcal{Q}}(\alpha)$ as an integrating factor for Eq.~\eqref{eq:dnstp}.
Operating from the left on the semi-discrete momentum equation of Eq.~\eqref{eq:dnstp} with $\mathsf{E}_{\mathcal{F}}\left(\frac{t-\tau}{(\Delta x)^2\text{Re}}\right)$ and introducing the transformed variable $\mathsf{v} = \mathsf{E}_{\mathcal{F}}\left(\frac{t-\tau}{(\Delta x)^2\text{Re}}\right) \mathsf{u}$ yields the transformed system of semi-discrete equations
\begin{equation}
  \frac{ d \mathsf{v} }{ d t }
    = -\mathsf{H}_{\mathcal{F}}
      \tilde{\mathsf{N}} \left( \mathsf{H}_{\mathcal{F}}^{-1} \mathsf{v} + \mathsf{u}_\infty \right)
      - \mathsf{H}_{\mathcal{F}}\mathsf{G} \mathsf{d},\quad
  \mathsf{G}^\dagger \mathsf{H}_{\mathcal{C}}^{-1} \mathsf{v} = 0,
  \label{eq:dnstp_trans_0}
\end{equation}
where $\mathsf{H}_\mathcal{Q} = \mathsf{E}_{\mathcal{Q}}\left(\frac{t-\tau}{(\Delta x)^2\text{Re}}\right)$.
Using the commutativity properties of integrating factors, Eq.~\eqref{eq:dnstp_trans_0} simplifies to
\begin{equation}
  \frac{ d \mathsf{v} }{ d t }
    = -\mathsf{H}_{\mathcal{F}} \tilde{\mathsf{N}} \left(
      \mathsf{H}_{\mathcal{F}}^{-1} \mathsf{v} + \mathsf{u}_\infty \right)
      - \mathsf{G} \mathsf{b},\quad
  \mathsf{G}^\dagger \mathsf{v} = 0,
  \label{eq:dnstp_trans}
\end{equation}
where $\mathsf{b} = \mathsf{H}_\mathcal{F} \mathsf{d}$.
We emphasize that the transformed system of equations Eq.~\eqref{eq:dnstp_trans} is equivalent to the original system of equation Eq.~\eqref{eq:dnstp}.
Furthermore, as is the case for Eq.~\eqref{eq:dnstp}, Eq.~\eqref{eq:dnstp_trans} represents a system of DAEs of index 2.

The procedures for obtaining expressions $\mathsf{G}_{\mathsf{L}}(\mathbf{n})$ can be readily extended to the case of $[\mathsf{G}_{\mathsf{E}}(\alpha)](\mathbf{n})$, where $\mathsf{G}_{\mathsf{E}}(\alpha)$ is the LGF of the integrating factor $\mathsf{E}_{\mathcal{Q}}(-\alpha)$.
Expressions for $\mathsf{G}_{\mathsf{E}}(\mathbf{n})$ are also provided in \ref{app:lgfs}.
As for the case of $\mathsf{L}_\mathcal{Q}^{-1}$, fast solutions to expressions involving $\mathsf{G}_{\mathsf{E}}(\alpha)$ are computed using the LGF-FMM.

An important distinction between $\mathsf{G}_{\mathsf{L}}(\mathbf{n})$ and $[\mathsf{G}_{\mathsf{E}}(\alpha)](\mathbf{n})$ is found in their asymptotic behavior.
Whereas $|\mathsf{G}_{\mathsf{L}}(\mathbf{n})|$ decays as $1/|\mathbf{n}|$ as $|\mathbf{n}|\rightarrow \infty$, $|[\mathsf{G}_{\mathsf{E}}(\alpha)](\mathbf{n})|$ decays faster than any exponential as $|\mathbf{n}|\rightarrow\infty$ for a fixed $\alpha$.%
\footnote{%
Consider $[\mathsf{G}_{\mathsf{E}}(\alpha)](\mathbf{n})$ for the case $\mathbf{n}=(n,0,0)$.
As $n\rightarrow\infty$, $[\mathsf{G}_{\mathsf{E}}(\alpha)](\mathbf{n}) \sim \alpha^n/n!$.
For $\alpha=0.1$ and $\alpha=1.0$, the value of $[\mathsf{G}_{\mathsf{E}}(\alpha)](\mathbf{n})/[\mathsf{G}_{\mathsf{E}}(\alpha)](\mathbf{0})$ is less than $10^{-10}$ at $n=7$ and $n=13$, respectively.
The numerical simulations of Section~\ref{sec:verif} make use of integrating factors with $\alpha<1$, but larger values of $\alpha$ are allowed.
}
The fast decay of $\mathsf{G}_{\mathsf{E}}$ implies that, for typical computations, the application of $\mathsf{E}_{\mathcal{Q}}$ can be consider a local operation, i.e. values computed at a particular grid location only depend on the values of a few neighboring grid cells.
Consequently, the LGF-FMM requires significantly fewer operations to evaluate the action of $\mathsf{E}_{\mathcal{Q}}$ compared to the action of $\mathsf{L}_{\mathcal{Q}}^{-1}$.\footnote{%
For the run parameters of the numerical experiments of Section~\ref{sec:verif}, the action of $\mathsf{E}_\mathcal{Q}$ only requires approximately 10\% of the total number of operations required to compute $\mathsf{L}_\mathcal{Q}^{-1}$.
}

\section{Time integration}
\label{sec:temporal}

\subsection{Half-explicit Runge-Kutta methods}
\label{sec:temporal_herk}

Failing to properly identify the semi-discrete form of the governing equations, i.e. Eq.~\eqref{eq:dnstp}, as a system of differential algebraic equations (DAEs) of index 2 prior to choosing a time integration scheme can have undesirable consequences on the quality of the numerical solution \cite{hairer1996,ascher1998}.
Half-explicit Runge-Kutta (HERK) methods are a type of one-step time integration schemes developed for DAEs of index 2 \cite{hairer1989,brasey1993,hairer1996}.
Although there are multiple HERK methods \cite{hairer1996}, we limit our attention to the original HERK method proposed by \citet{hairer1989}.

Consider DAE systems of index 2 of the form
\begin{equation}
  \frac{dy}{dt} = f \left( y, z \right),\quad g\left( y \right) = 0,
  \label{eq:dae_brasey}
\end{equation}
where $f$ and $g$ are sufficiently differentiable, and $z$ is an unknown that must be computed so as to have $y$ satisfy $g(y) = 0$.
Problems of this form are of index 2 if the product of partial derivatives $g_y(y) f_z(y,z)$ is non-singular in a neighborhood of the solution.
The HERK method applied to Eq.~\eqref{eq:dae_brasey} is given by an algorithm similar to that of explicit Runge-Kutta (ERK) methods except that the implicit constraint equation $g\left( y \right) = 0$ is solved at each stage of the ERK scheme.

Similarly to standard RK methods, HERK methods can be described by their Butcher tableau:
\begin{equation}
  \begin{array}{c|c}
    \mathbf{c} & \mathbf{A} \\ \hline
    {} & \mathbf{b}^\dagger
  \end{array},
  \label{eq:erk_tableau}
\end{equation}
where $\mathbf{A}=[a_{i,j}]$ is the Runge-Kutta matrix, $\mathbf{b}=[b_i]$ is the weight vector, and $\mathbf{c}=[c_i]$ is the node vector.
In subsequent sections, it is often convenient to use the \emph{shifted} tableau notation:
\begin{equation}
  \tilde{a}_{i,j} =
  \left\{\begin{array}{cl}
    a_{i+1,j} & \text{for } i = 1,2,\dots,s-1\\
    b_j & \text{for } i = s
  \end{array}\right.,\,\,
  \tilde{c}_i =
  \left\{\begin{array}{cl}
      c_{i+1} & \text{for } i = 1,2,\dots,s-1\\
      1 & \text{for } i = s
  \end{array}\right..
  \label{eq:herk_shift}
\end{equation}
We refer the reader to the discussions of \cite{hairer1989,brasey1993} for a detailed algorithm and a list of order-conditions for the general case of Eq.~\eqref{eq:dae_brasey}.

We now turn our attention to the special case of the transformed semi-discrete governing equations given by Eq.~\eqref{eq:dnstp_trans}.
It is convenient to express the non-autonomous system of Eq.~\eqref{eq:dnstp_trans} in terms of the autonomous system of Eq.~\eqref{eq:dae_brasey}.
This is achieved by letting $y=[ \mathsf{v}, t ]$ and $z = \mathsf{b}$, and by adding $t^\prime=1$ to Eq.~\eqref{eq:dnstp_trans}.
For this case, $g_y = [\,\mathsf{G}^\dagger,\, 0]$ and $f_z = g_y^\dagger$, where $g_y = [ g_\mathsf{u}, g_t ]$ and $f_z=f_\mathsf{b}$.
By construction, the operator $\mathsf{G}$ is a constant, which implies that $f_z$ and $g_y$ are also constants.
As a result, order-conditions for the general system of Eq.~\eqref{eq:dae_brasey} involving high-order derivatives of $f_z$ and $g_y$ are trivially satisfied for the case of Eq.~\eqref{eq:dnstp_trans}.
Fewer order-conditions permit a wider range of RK tableaus to be used for a given order of accuracy.
This is particularly relevant for high-order HERK schemes, since the number of order-conditions is significantly larger than that of standard RK schemes \cite{brasey1993}.

The simplifications in the order-conditions obtained for the special case of constant $f_z$ and $g_y$ are well-described in the literature of HERK methods \cite{hairer1989,brasey1993,hairer1996,sanderse2012}.
Order-conditions up to order 4 for the $y$-component reduce to those of standard RK methods \cite{sanderse2012}.
Similarly, order-conditions of order $r \le 3$ for the $z$-component (up to fourth-order accurate $z$-component) reduce to having the shifted sub-tableau $[\tilde{a}_{i,j}]$ for $i,j=1,2,\dots s-1$ satisfy the $y$-component order-conditions up to order $r$ \cite{brasey1993,sanderse2012}.
It is beyond the scope of the present work to provide an extended discussion on the properties and implementation details of the HERK method for particular RK tableaus.
Instead, the order of accuracy and linear stability of a few selected schemes used to perform the numerical experiments of Section~\ref{sec:verif} is discussed in Section~\ref{sec:temporal_ifherk} and \ref{app:stability}, respectively.

\subsection{Combined integrating factor and half-explicit Runge-Kutta method}
\label{sec:temporal_ifherk}

In this section we present a method for obtaining numerical solutions for the (untransformed) discrete velocity and total pressure perturbation by combining the integrating factor technique of Section~\ref{sec:spatial_intfact} with the HERK method of Section~\ref{sec:temporal_herk}.
The combined method, referred to as the IF-HERK method, integrates Eq.~\eqref{eq:dns} over $t\in[0,T]$ subject to the initial condition $\mathsf{u}(\mathbf{n},0) = \mathsf{u}_0(\mathbf{n})$.

Formally, the IF-HERK method partitions the original problem into a sequences of $n$ sub-problems, where the $k$-th sub-problem corresponds to numerical integration of Eq.~\eqref{eq:dns} from $t_k$ to $t_{k+1}$ subject to the initial condition $\mathsf{u}(\mathbf{n},t_k) = \mathsf{u}_k(\mathbf{n})$.
We restrict our discussion to the case of equispaced time-steps, i.e. $t_k = t_{k-1} + \Delta t$, since the more general case of variable time-step size is readily deduced.

The $k$-th sub-problem is solved by first introducing the transformed variables
\begin{equation}
  \mathsf{v}(\mathbf{n},t) = \left[ \mathsf{E}_\mathcal{F}
     \left(\textstyle\frac{\Delta t}{(\Delta x)^2\text{Re}}\right) \right]
      \mathsf{u}(\mathbf{n},t), \,\,
  \mathsf{b}(\mathbf{n},t) = \left[ \mathsf{E}_\mathcal{F}
    \left(\textstyle\frac{\Delta t}{(\Delta x)^2\text{Re}}\right) \right]
      \mathsf{q}(\mathbf{n},t), \,\,
    t \in[t_k,t_{k+1}],
  \label{eq:ifherk_aux}
\end{equation}
and using $\mathsf{E}_\mathcal{F}\left(\textstyle\frac{\Delta t}{(\Delta x)^2\text{Re}}\right)$ as an integrating factor for Eq.~\eqref{eq:dnstp}.
Next, the HERK method is used to integrate the transformed nonlinear equations from $t_k$ to $t_{k+1}$ in order to obtain $\mathsf{v}_{k+1}(\mathbf{n}) \approx \mathsf{v}(\mathbf{n},t_{k+1})$ and $\mathsf{b}_{k+1} \approx \mathsf{b}(\mathbf{n},t_{k+1})$.
Finally, values for the discrete velocity and total pressure perturbation at $t_{k+1}$, i.e. $\mathsf{u}_{k+1}(\mathbf{n}) \approx \mathsf{u}(\mathbf{n},t_{k+1})$ and $\mathsf{d}_{k+1}(\mathbf{n}) \approx \mathsf{d}(\mathbf{n},t_{k+1})$, are obtained from $\mathsf{v}_{k+1}$ and $\mathsf{b}_{k+1}$ by using the integrating factor $\mathsf{E}_\mathcal{F}\left(\textstyle\frac{-\Delta t}{(\Delta x)^2\text{Re}}\right)$.

A computationally convenient algorithm for the $k$-th time-step of the IF-HERK method, subsequently denoted by $(\mathsf{u}_{k+1},{t}_{k+1},\mathsf{p}_{k+1}) \leftarrow \text{IF-HERK}(\mathsf{u}_{k},t_k)$, is given by:
\begin{enumerate}
\item \emph{initialize}: copy solution values from the $k$-th time-step,
  \begin{equation}
    \mathsf{u}_k^0 = \mathsf{u}_k, \quad t^{0}_{k} = t_k.
  \end{equation}
\item \emph{multi-stage}: for $i=1,2,\dots,s$, solve the linear system
  \begin{equation}
    \left[ \begin{array}{cc}
      \left( \mathsf{H}_\mathcal{F}^{i} \right)^{-1} & \mathsf{G} \\
      \mathsf{G}^\dagger & 0
    \end{array} \right]
    \left[ \begin{array}{c}
      \mathsf{u}_k^i \\
      \hat{\mathsf{d}}_k^i
    \end{array} \right]
    =
    \left[ \begin{array}{c}
      \mathsf{r}_k^i \\
      0
    \end{array} \right],
    \label{eq:ifherk_linsys}
  \end{equation}
  where
  \begin{equation}
    \mathsf{H}_\mathcal{F}^i =
      \mathsf{E}_\mathcal{F}\left(\textstyle\frac{ (\tilde{c}_i-\tilde{c}_{i-1}) \Delta t}{(\Delta x)^2\text{Re}}\right),
    \quad
    \mathsf{r}_k^i = \mathsf{q}_k^{i}
      + \Delta t \sum_{j=1}^{i-1} \tilde{a}_{i,j} \mathsf{w}_k^{i,j}
      + \mathsf{g}_k^{i},
    \label{eq:ifherk_aux_1}
  \end{equation}%
  \begin{equation}
    \mathsf{g}_k^i = \
      - \tilde{a}_{i,i} \Delta t \
      \tilde{\mathsf{N}}\left( \mathsf{u}_k^{i-1} + \mathsf{u}_\infty(t_k^{i-1})\right),
      \quad
      t_k^{i} = t_k + \tilde{c}_i \Delta t.
    \label{eq:ifherk_g}%
  \end{equation}
  For $i>1$ and $j>i$, $\mathsf{q}_k^{i}$ and $\mathsf{w}_k^{i,j}$ are recursively computed using%
  \footnote{%
  An efficient implementation of the IF-HERK algorithm recognizes that the application of $s-1$ integrating factors can be avoided during final, $i=s$, stage by computing $\mathsf{r}_k^s = \mathsf{H}_\mathcal{F}^{i-1} \left( \mathsf{q}_k^{s-1} + \Delta t \sum_{j=1}^{i-1} \tilde{a}_{i,j} \mathsf{w}_k^{i-1,j}\right) + \mathsf{g}_k^{i}$, as opposed to Eq.~\eqref{eq:ifherk_g}.
  This modification avoids having to explicitly compute $\mathsf{q}_k^{s}$ and $\mathsf{w}_k^{s,j}$ for $j=1,2,\dots s-1$.%
  }
  \begin{equation}
    \mathsf{q}_k^{i} = \mathsf{H}_\mathcal{F}^{i-1} \mathsf{q}_k^{i-1},
    \quad
    \mathsf{q}_k^{1} = \mathsf{u}_k^0
    \label{eq:ifherk_q}
  \end{equation}%
  \begin{equation}
    \mathsf{w}_k^{i,j} = \mathsf{H}_\mathcal{F}^{i-1} \mathsf{w}_k^{i-1,j},
    \quad
    \mathsf{w}_k^{i,i} = \left( \tilde{a}_{i,i} \Delta t \right)^{-1}
      \left( \mathsf{g}_k^{i} - \mathsf{G} \hat{\mathsf{d}}_k^{i} \right).
  \end{equation}
\item \emph{finalize}: define the solution and constraint values of the $(k+1)$-th time-step,
  \begin{equation}
    \mathsf{u}_{k+1} = \mathsf{u}_k^s,
    \quad
    \mathsf{d}_{k+1} = \left( \tilde{a}_{s,s} \Delta t \right)^{-1} \hat{\mathsf{d}}_k^s,
    \quad
    t_{k+1} = t_k^{s}.
  \end{equation}%
\end{enumerate}
The above algorithm is obtained by applying the HERK method to either Eq.~\eqref{eq:dnstp_trans} or, equivalently, Eq.~\eqref{eq:dnstp_trans_0} for the $k$-th sub-problem, and introducing the auxiliary variables
\begin{equation}
  \mathsf{u}^{i}_{k}(\mathbf{n})
    = \left[ \mathsf{E}_\mathcal{F}
      \left(\textstyle\frac{-\tilde{c}_i \Delta t}{(\Delta x)^2\text{Re}}\right) \right]
      \mathsf{v}^{i}_{k} (\mathbf{n}),
  \quad
  \mathsf{d}^{i}_{k}(\mathbf{n})
    = \left[ \mathsf{E}_\mathcal{F}
      \left(\textstyle\frac{-\tilde{c}_i \Delta t}{(\Delta x)^2\text{Re}}\right) \right]
      \mathsf{b}^{i}_{k}(\mathbf{n}),
\end{equation}
for $i= 1, 2, \dots s$.
We clarify that the intermediate steps used to obtained the final form $\text{IF-HERK}$ algorithm make use of the commutativity properties of $\mathsf{E}_\mathcal{Q}$ and the identity $\mathsf{E}_\mathcal{Q}(\alpha_1) \mathsf{E}_\mathcal{Q}(\alpha_2) = \mathsf{E}_\mathcal{Q}(\alpha_1+\alpha_2)$.

The linear operator on the left-hand-side (LHS) of Eq.~\eqref{eq:ifherk_linsys} is symmetric positive semi-definite and its null-space is spanned by the set of $[ 0, \mathsf{a} ]^\dagger$, where $\mathsf{a}\in\mathbb{R}^\mathcal{C}\times\mathbb{R}$ is any discrete linear polynomial.
Consequently, the compatibility condition on the pressure field given by Eq.~\eqref{eq:press_compat} guarantees Eq.~\eqref{eq:ifherk_linsys} has a unique solution.
As presented, the $\text{IF-HERK}$ algorithm is compatible with any HERK scheme since no assumptions have been made on the RK coefficients.
Of course, more efficient versions of this algorithm can potentially be obtained for specific families of RK coefficients, but such details are beyond the scope of the present work.

The IF-HERK schemes used to performed the numerical experiments of Section~\ref{sec:verif} are given by the following tableaus:
\begin{equation}
  \stackrel{\mbox{\small{\textsc{Scheme A}}}}{%
    \begin{array}{c|ccc}
      0 & 0 & 0 & 0 \\
      \textstyle\frac{1}{2} & \textstyle\frac{1}{2} & 0 & 0 \\
      1 & \textstyle\frac{\sqrt{3}}{3} & \textstyle\frac{3-\sqrt{3}}{3} & 0 \\
      \hline
      {} & \textstyle\frac{3+\sqrt{3}}{6} & -\textstyle\frac{\sqrt{3}}{3} & \textstyle\frac{3+\sqrt{3}}{6}
    \end{array} }, \quad
  \stackrel{\mbox{\small{\textsc{Scheme B}}}}{%
  \begin{array}{c|ccc}
    0 & 0 & 0 & 0 \\
    \textstyle\frac{1}{3} & \textstyle\frac{1}{3} & 0 & 0 \\
    1 & -1 & 2 & 0 \\
    \hline
    {} & 0 & \textstyle\frac{3}{4} & \textstyle\frac{1}{4}
  \end{array} }, \quad
  \stackrel{\mbox{\small{\textsc{Scheme C}}}}{%
  \begin{array}{c|ccc}
    0 & 0 & 0 & 0 \\
    \textstyle\frac{8}{15} & \textstyle\frac{8}{15} & 0 & 0 \\
    \textstyle\frac{2}{3} & \textstyle\frac{1}{4} & \textstyle\frac{5}{12} & 0 \\
    \hline
    {} & \textstyle\frac{1}{4} & 0 & \textstyle\frac{3}{4}
  \end{array} }.
  \label{eq:ifherk_schemes}
\end{equation}
The order of accuracy, based on the simplified order-conditions discussed in Section~\ref{sec:temporal_herk}, for each scheme is provided in Table~\ref{tab:ifherk_schemes}.
As a point of comparison, Table~\ref{tab:ifherk_schemes} also provides the expected order of accuracy for general semi-explicit DAEs of index 2, i.e. Eq.~\eqref{eq:dae_brasey}.

\begin{table}[ht]
  \centering
  \caption{%
  Order of accuracy of the solution $y$ variable (velocity perturbation) and constraint $z$ variable (pressure perturbation) based on specialized HERK order conditions.
  The superscript $*$ denotes values for general semi-explicit DAEs of index 2.
  \label{tab:ifherk_schemes}
  }
  \begin{tabular}{c|ccccc}
    {} & \emph{$y$-Order} & \emph{$z$-Order} & \emph{$y$-Order}${}^*$ & \emph{$z$-Order}${}^*$ & \\
    \hline
    Scheme A & 2 & 2 & 2 & 2 \\
    Scheme B & 3 & 2 & 3 & 2 \\
    Scheme C & 3 & 1 & 2 & 1
  \end{tabular}
\end{table}

The tableaus for Schemes B and C were obtained from \cite{brasey1993} and \cite{sanderse2012}.
As discussed in \cite{sanderse2012}, the tableau for Scheme C corresponds to the RK coefficients of the popular three-stage fractional step method of \cite{le1991}.
Unlike Schemes B and C, the tableau for Scheme A was specifically defined for the IF-HERK method.
An advantage of Scheme A over Schemes B and C is that the RK nodes, $c_i$'s, are equally spaced.
As a result, the IF-HERK method only requires a single non-trivial integrating factor.%
\footnote{%
One additional integrating factor is required during the last stage of the IF-HERK algorithm, but for the case of $c_s=1$ this additional integrating factor reduces to the identify operator.
}
This reduction in the number of distinct LGFs reduces the number of pre-processing operations and lowers the storage requirements of the LGF-FMM.
Additionally, extensions of the present method including immersed surfaces, e.g. via the treatment of immersed boundaries of \cite{colonius2008}, can potentially enjoy similar reductions in the computational costs of pre-processing operations by only having to consider a single non-trivial integrating factor.
We will report on immersed boundary methods based on the present flow solver in subsequent publications.
The linear stability analysis of the IF-HERK method is provided in \ref{app:stability}.

\subsection{Projection method}
\label{sec:temporal_projection}

It is readily verified that the most computationally expensive operation performed by the IF-HERK method corresponds to solving Eq.~\eqref{eq:ifherk_linsys} for each stage.
Systems of continuum or discrete equations similar to Eq.~\eqref{eq:ifherk_linsys} often arise in the literature of numerical methods for simulating incompressible flows.
Solutions to these system are frequently obtained through classical projection, fractional-step, or pressure Schur complement methods \cite{perot1993,turek1999}.
These methods can be regarded as approximate block-wise LU decompositions of the original system \cite{perot1993,turek1999}.
More recently, \emph{exact} projection techniques that are free of any matrix/operator approximations have been proposed, e.g. \cite{chang2002,colonius2008}.
These techniques have the advantage of not introducing any ``splitting errors'' and do not require artificial pressure boundary conditions.
The present formulation uses an exact projection method to solve Eq.~\eqref{eq:ifherk_linsys}, but differs from the methods of \cite{chang2002,colonius2008} in that it does not use the null-space of the discrete operators to obtain solutions to the linear system.

The block-wise LU decomposition of the operator in Eq.~\eqref{eq:ifherk_linsys} suggests a solution procedure, expressed in the standard correction form, given by:
\begin{subequations}
  \begin{alignat}{2}
    \mathsf{u}^{*} 
      &= \mathsf{H}_{\mathcal{F}}^{i}  \mathsf{r}_k^{i}
      && \quad\text{(compute intermediate velocity)} \\
    \mathsf{S} \hat{\mathsf{d}}_k^{i} 
      &= \mathsf{G}^\dagger \mathsf{u}^{*}
      && \quad\text{(solve for total pressure)}
    \label{eq:proj_schur} \\
    \mathsf{u}_k^{i}
      &= \mathsf{u}^{*} - \mathsf{H}_{\mathcal{F}}^{i} \mathsf{G} \hat{\mathsf{d}}_k^{i}
      && \quad\text{(projection step)},%
  \end{alignat}
  \label{eq:proj}%
\end{subequations}
where $\mathsf{S} = \mathsf{G}^{\dagger} \mathsf{H}_\mathcal{F}^{i} \mathsf{G}$ is the Schur complement of the system.%
\footnote{%
Without additional information the (scaled) total pressure perturbation, $\hat{\mathsf{d}}_k^{i}$, obtained from Eq.~\eqref{eq:proj_schur} is unique up to a discrete linear polynomial.
Yet, a unique $\hat{\mathsf{d}}_k^{i}$ is obtained by taking into account the compatibility condition $\mathsf{p}(\mathbf{n},t) \rightarrow 0$ as $|\mathbf{n}| \rightarrow \infty$, i.e. $\hat{\mathsf{d}}_k^{i}(\mathbf{n}) \rightarrow \mathsf{c}^i_k$ as $|\mathbf{n}| \rightarrow \infty$ where $\mathsf{c}^i_k = \frac{1}{2} |\mathsf{u}_\infty(t^i_k)|^2$, discussed in Section~\ref{sec:spatial_discrete}.%
}
By taking into account the commutativity and mimetic properties of the spatial discretization scheme the procedure given by Eq.~\eqref{eq:proj} simplifies to:
\begin{equation}
  \hat{\mathsf{d}}_k^{i} = - \mathsf{L}_\mathcal{C}^{-1} \mathsf{G}^\dagger \mathsf{r}_k^{i},
  \quad
  \mathsf{u}_k^{i} = \mathsf{H}_{\mathcal{F}}^{i} \left( \mathsf{r}_k^{i} - \mathsf{G} \hat{\mathsf{d}}_k^{i} \right),
  \label{eq:proj_final}
\end{equation}
where $\mathsf{x} = \mathsf{L}_\mathcal{C}^{-1} \mathsf{y}$ is equivalent to solving $\mathsf{L}_\mathcal{C} \mathsf{x} = \mathsf{y}$ subject to uniform boundary conditions at infinity.
In this form, one of the two integrating factors has been eliminated and the original elliptic problem $\mathsf{G}^{\dagger} \mathsf{H}_\mathcal{F}^{i} \mathsf{G} \mathsf{x} = \mathsf{y}$ has been replaced by the Poisson problem $\mathsf{L} \mathsf{x} = \mathsf{y}$.
Reducing the original discrete elliptic problem to a discrete Poisson problem is of significant practical importance since it permits the use of the LGF-FMM with known LGF expressions \cite{liska2014}.
As will be discussed in Section~\ref{sec:truncation}, the operation count of our overall algorithm is dominated by the cost of solving for the discrete pressure perturbation; therefore, a projection method that is compatible with fast, robust discrete elliptic solvers greatly facilitates obtaining fast flow solutions.

\section{Adaptive computational grid}
\label{sec:truncation}

\subsection{Restricting operations to a finite computational grid}
\label{sec:truncation_active}

Thus far we have described algorithms for discretizing and computing the incompressible Navier-Stokes equations on unbounded grids.
In this section, we present a method for computing solutions, to a prescribed tolerance, using only a finite number of operations.
This approximation is accomplished by limiting all operations to a finite computational grid obtained by removing grid cells of the original unbounded grid containing field values that are sufficiently small so as not to significantly affect the evolution of the flow field.
As will be demonstrated in the following discussion, the ability of the present method to only track a finite region of the unbounded domain is a consequence of the exponential decay of the vorticity at large distances, which is assumed for all flows under consideration.

We first consider the error resulting from neglecting field values outside a finite region when solving the elliptic problems of the IF-HERK method.%
\footnote{%
Field values outside the finite region being tracked are treated as zero.
}
Using the notation of Section~\ref{sec:spatial_lgfs}, the solution to the discrete Poisson problem of Eq.~\eqref{eq:proj_final} is given by
\begin{equation}
  \hat{\mathsf{d}}(\mathbf{n}) = [ \mathsf{G}_\mathsf{L} * \mathsf{f} ](\mathbf{n}),
  \quad \
  \mathsf{f}(\mathbf{n}) = [-\mathsf{G}^\dagger \mathsf{r}_k^{i}](\mathbf{n}).
  \label{eq:dpoisson_conv_d}
\end{equation}
The source field $\mathsf{G}^\dagger \mathsf{r}_k^{i}$ is a discrete approximation of $\nabla \cdot \boldsymbol{\ell}$ at $t \approx k \Delta t$, where $\boldsymbol{\ell} = \boldsymbol{\omega} \times \mathbf{u}$ is the Lamb vector.
It follows from the assumption that $\boldsymbol{\omega}$ is exponentially small at large distances that $\nabla \cdot \boldsymbol{\ell}$ and $\mathsf{G}^\dagger \mathsf{r}_k^{i}$ must also be exponentially small at large distances.
As a result, the induced field of Eq.~\eqref{eq:dpoisson_conv_d} is computed to a prescribed tolerance by defining the finite computational domain such that it includes the region where the magnitude of $\mathsf{G}^\dagger \mathsf{r}_k^{i}$ is greater than some positive value.

The action of all operators present in the IF-HERK and projection algorithms, with the exception of $\mathsf{L}_\mathcal{C}^{-1}$, are evaluated using only a few local operations.
Many of these local operators act on fields that typically decay algebraically, e.g. $\mathsf{u}$ and $\mathsf{d}$.
As a result, the technique of only tracking regions with non-negligible source terms used for Eq.~\eqref{eq:proj_final} is impractical for most other operations required by the IF-HERK method.
Unlike the action of $\mathsf{L}_\mathcal{C}^{-1}$, the action of local operators only incurs an error limited to a few cells near the boundary of a finite region if field values of outside the region are ignored, i.e. taken to be zero.
Furthermore, repeated applications of local operators only propagate the error into the interior of the region by a few grid cells per application.
This type of error is prevented from significantly affecting the solution in the interior by padding the interior with buffer grid cells and by periodically computing (``refreshing'') $\mathsf{u}$ from the discrete vorticity, $\mathsf{w} = \mathsf{C}\mathsf{u}$, which, like $\mathsf{G}^\dagger \mathsf{r}_k^{i}$, has bounded approximate support.
As a result, the approximate support of both $\mathsf{G}^\dagger \mathsf{r}^i_k$ and $\mathsf{w}$ must be contained in the finite computational domain.
Bounds for the error resulting from approximating the support of these fields and estimates for the number of time steps that can elapse before the velocity needs to refreshed will be discussed in Sections~\ref{sec:truncation_adaptivity} and \ref{sec:truncation_refresh}, respectively.

We recall that the discrete velocity perturbation $\mathsf{u}$ is subject to the constraint $\mathsf{G}^\dagger\mathsf{u}=0$ and that the null-space of $\mathsf{G}^\dagger$ is spanned by the image of $\mathsf{C}^\dagger$.
As a result, it is possible to express $\mathsf{u}$ as
\begin{equation}
  \mathsf{u} = \mathsf{C}^\dagger \mathsf{a},
  \label{eq:vel_stf}
\end{equation}
where $\mathsf{a}\in\mathbb{R}^\mathcal{E}$ can be regarded as the discrete vector potential or streamfunction.
Additionally, we require $\mathsf{D}\mathsf{a}=0$.
The discrete vorticity, $\mathsf{w}$, can now be expressed in terms of $\mathsf{a}$ as
\begin{equation}
  \mathsf{w}
    = \mathsf{C} \mathsf{C}^\dagger \mathsf{a}
    = \left( \mathsf{C} \mathsf{C}^\dagger + \mathsf{D}^\dagger \mathsf{D}\right) \mathsf{a}
    = -\mathsf{L}_\mathcal{E} \mathsf{a}.
  \quad \mathsf{D} \mathsf{w} = 0
    \label{eq:vor_stf}
\end{equation}
Finally, Eq.~\eqref{eq:vel_stf} and \eqref{eq:vor_stf} provide an expression for $\mathsf{u}$ in terms of $\mathsf{w}$,
\begin{equation}
  \mathsf{u} = -\mathsf{C}^\dagger \mathsf{L}_\mathcal{E}^{-1} \mathsf{w},
  \label{eq:vel_vor}
\end{equation}
where $\mathsf{L}_\mathcal{E}^{-1}$ imposes zero boundary conditions at infinity.%
\footnote{%
Without further considerations Eq.~\eqref{eq:vor_stf} implies that $\mathsf{a}$ is unique up to a discrete linear polynomial. Given that $\mathsf{w}$ is exponentially small at large distances and that $\mathsf{u}$ tends to zero at infinity, it follows that $\mathsf{a}$ is unique up to an arbitrary constant taken to be zero.
}
As expected, the expressions relating $\mathsf{u}$, $\mathsf{w}$, and $\mathsf{a}$ are analogous to the continuum expressions relating the velocity, vorticity, and streamfunction fields.
We emphasize that Eq.~\eqref{eq:vel_stf}, \eqref{eq:vor_stf}, and \eqref{eq:vel_vor} were obtained through the algebraic properties of the discrete operators, as opposed to the discretization of continuum equations.

The present formulation can be cast into an equivalent vorticity formulation simply by taking the discrete curl of Eq.~\eqref{eq:dnstp} and computing $\mathsf{u}$, which is required to evaluate the non-linear term, using Eq.~\eqref{eq:vel_vor}.
This formulation is not pursued since each stage of the IF-HERK would require solving a discrete \emph{vector} Poisson problem, as opposed to a discrete \emph{scalar} Poisson problem, which would in turn roughly triple the cost of each stage.%
\footnote{%
For the test case of the extremely thin $\delta/R = 0.0125$ vortex ring discussed in Section~\ref{sec:verif-selfindc}, the wall-time ratio of a vector to a scalar discrete Poisson solve is approximately 2.8, which is slightly less than the expected ratio of 3 based on operation count estimates of the LGF-FMM due to the larger parallel communication costs per problem unknown for the scalar case.%
}
The vorticity formulation has the advantage of not having to periodically evaluate Eq.~\eqref{eq:vel_vor} to refresh $\mathsf{u}$, but, as will be discussed in Sections~\ref{sec:truncation_refresh}, this operation occurs, at most, once per time step.
Based on the stability analysis of \ref{app:stability}, RK schemes with a minimum of three stages are required to ensure stable solutions.
As a result, the primitive variable formulation is approximately 1.5 to 3 times faster than the vorticity formulation.
Differences in the errors between the two algebraically-equivalent formulations associated with the finite tolerances used to compute the LGF-FMM and the adaptive grid algorithms can be used to further distinguish each formulation.
However, such differences in errors are not considered here since they are expected to be on the order of the prescribed tolerances, which, as will be discussed in Section~\ref{sec:algorithm}, are specified to be much smaller than the discretization errors of practical flows.

\subsection{Block-structured active computational grid}
\label{sec:truncation_blocks}

We now turn our attention to the formal definition of the finite region of the unbounded computational domain tracked by our formulation, which we refer to as the \emph{active} computational domain.
Consider partitioning the unbounded staggered Cartesian grid described in Section~\ref{sec:spatial} into an infinite set of equally sized blocks arranged on a logically Cartesian grid.
The block corresponding to the $\mathbf{n}=(i,j,k)$ location is denoted by $B(\mathbf{n})$ or, equivalently, $B_{i,j,k}$, and the union of all blocks is denoted by $D_\infty$.
Each block is defined as a finite staggered Cartesian grid of $n^b_{1} \times n^b_{2} \times n^b_{3}$ cells.
We limit our attention to the case in which each block contains the same number of cells in each direction, i.e. $n^b_i = n^b$, but note that the subsequent discussion readily extends to the general case.
As a practical consideration, a layer of buffer or ghost grid cells surrounding each block is introduced to facilitate the implementation of the present algorithm.

\begin{figure}[htbp]
  \begin{center}
    \includegraphics[width=\textwidth]{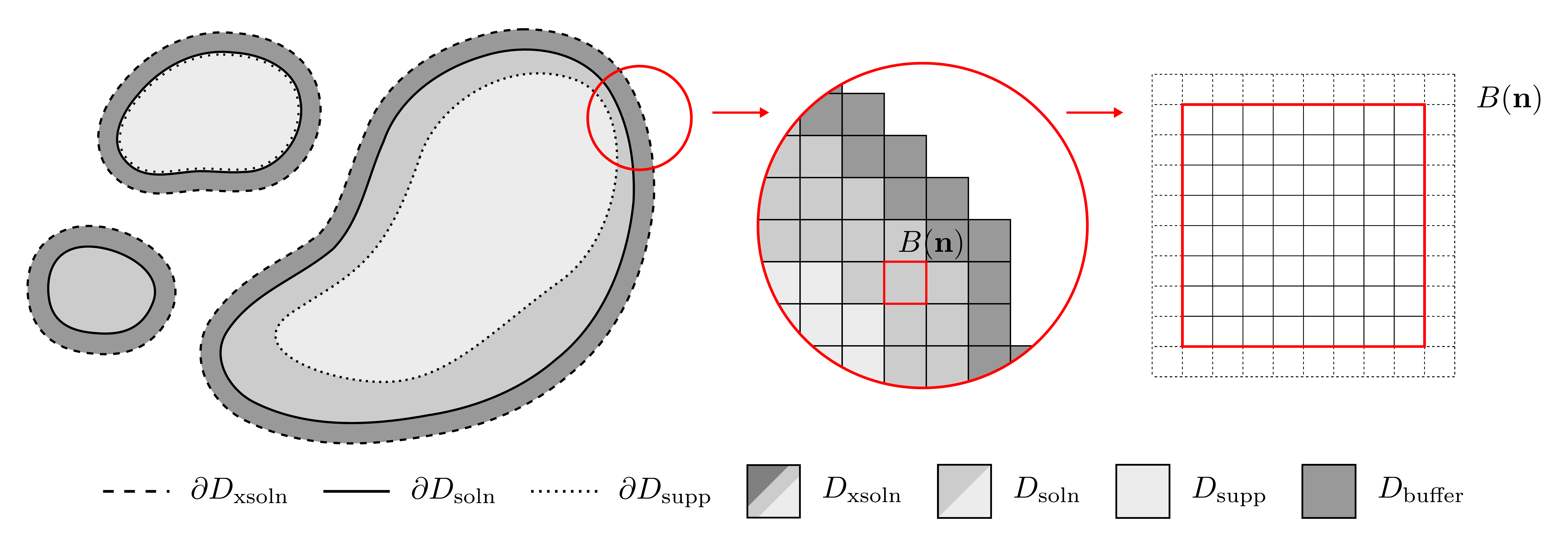}
  \end{center}
  \caption{%
  Depiction of the finite computational domain in two-dimensions.
  Distant view of the three nested sub-domains $D_\text{supp} \subseteq D_\text{soln} \subset D_\text{xsoln}$ defined in the main text (\emph{left}).
  Zoomed-in view illustrating the union of blocks used to define the domain (\emph{middle}).
  Magnified view of an individual block (\emph{right}).
  Each block is defined as a finite staggered Cartesian grid; dashed cells surrounding the interior grid correspond to buffer or ghost grid cells.
  \label{fig:grid-blocks}
  }
\end{figure}

Figure~\ref{fig:grid-blocks} depicts the three nested sub-domains $D_\text{supp} \subseteq D_\text{soln} \subset D_\text{xsoln} \subset D_\infty$ that constitute the active computational domain.
These sub-domains are defined as:
\begin{itemize}
  \item \emph{Support blocks} ($D_\text{supp}$): union of blocks that defines the support of the source field of the discrete Poisson problems of Eq.~\eqref{eq:proj_final} and \eqref{eq:vor_stf}.
  \item \emph{Solution blocks} ($D_\text{soln}$): union of blocks that tracks the solution fields $\mathsf{u}$ and $\mathsf{d}$.
  All field values defined in the blocks belonging to $D_\text{soln}$ are regarded as accurate approximations of the field values computed using an unbounded domain.
  \item \emph{Expanded solution blocks} ($D_\text{xsoln}$): union of blocks given by a non-trivial neighborhood of $D_\text{soln}$.
  We limit our attention to neighborhoods defined by the union of blocks that are at most $N_b$ blocks away from any block contained in $D_\text{soln}$,
  \begin{equation}
    D_\text{xsoln} = \left\{ B(\mathbf{m}) :
      | \mathbf{n}-\mathbf{m} | \le N_b,\,\,
        B(\mathbf{n}) \in D_\text{soln},\,\,
        \mathbf{m},\mathbf{n} \in \mathbb{Z}^3 \right\}.
    \label{eq:def_xsoln}
  \end{equation}
  \item \emph{Buffer blocks} ($D_\text{buffer}$): union of blocks belonging to $D_\text{xsoln}$, but not belonging to $D_\text{soln}$, i.e. $D_\text{buffer}=D_\text{xsoln} \setminus D_\text{soln}$.
  (The domain $D_\text{buffer}$ is not one of the three primary sub-domains, but it is introduced to facilitate the subsequent discussion.)
\end{itemize}
The criteria for selecting which blocks belong to $D_\text{supp}$ and $D_\text{soln}$ are discussed in Section~\ref{sec:truncation_adaptivity}, and the techniques for selecting values of $N_n$ discussed in Section~\ref{sec:truncation_refresh}.

We now introduce the ``mask operator'' $\mathsf{M}^{\gamma}_\mathcal{Q}:\mathbb{R}^\mathcal{Q}\mapsto\mathbb{R}^\mathcal{Q}$ associated with the grid space $\mathcal{Q}$ and the domain $\gamma$, which is defined by
\begin{equation}
  [ \mathsf{M}^{\gamma}_\mathcal{Q} \mathsf{q} ] (\mathbf{n}) =
  \left\{
    \begin{array}{cc}
      \mathsf{q}(\mathbf{n}) &
        \,\, \text{if} \,\, \mathbf{n} \in \text{\textsl{ind}}[B]
        \,\, \text{and} \,\, B \in D_\gamma \\
      0 & \text{otherwise}
    \end{array}
  \right.,
\end{equation}
where $\mathsf{q}\in\mathbb{R}^\mathcal{Q}$, and $\text{\textsl{ind}}[B]$ denotes the set of all indices of the unbounded staggered grid associated with block $B$.
Mask operators are subsequently used to formally define operations performed on finite domains.
For example, the operation $\mathsf{G} \mathsf{d}$ perform over $D_\text{xsoln}$ is defined as $\mathsf{M}^{\text{xsoln}}_\mathcal{F} \mathsf{G} \mathsf{M}^{\text{xsoln}}_\mathcal{C} \mathsf{d}$.
For this particular operation, the values of $\mathsf{M}^{\text{xsoln}}_\mathcal{C} \mathsf{G} \mathsf{M}^{\text{xsoln}}_\mathcal{C} \mathsf{d}$ and $\mathsf{G} \mathsf{d}$ are equivalent for grid cells in $D_\text{xsoln}$, except for a single layer of grid cells on the boundary of $D_\text{xsoln}$.
Computationally efficient implementations of $\mathsf{M}^{\gamma^\prime}_{\mathcal{Q}^\prime}\mathsf{A}\mathsf{M}^{\gamma}_\mathcal{Q}$ recognize that all non-trivial numerical operations are limited to grid cells contained in either $D_\gamma$ and $D_{\gamma^\prime}$.

\subsection{Adaptivity}
\label{sec:truncation_adaptivity}

In this section we discuss the criteria used to select the blocks belonging to $D_\text{supp}$ and $D_\text{soln}$.
It follows from subsequent discussions that the field values on $D_\text{soln} \setminus D_\text{supp}$ can be computed as a post-processing step from the field values on $D_\text{supp}$; therefore, only the criteria used to define the $D_\text{supp}$ affects the accuracy of the computed flow field.
We allow for $D_\text{soln} \ne D_\text{supp}$ in order to emphasize that the present algorithm is able to track values of $\mathsf{u}$ and $\mathsf{d}$ over arbitrary regions of interest.

Consider a function $W$ that maps an unbounded grid of blocks, i.e. $D_\infty$, to an unbounded grid of positive real scalars.
We define the support and solution regions as
\begin{subequations}
  \begin{align}
    D_\text{supp} &= \left\{ B(\mathbf{n}) : [ W_\text{supp}(D_\infty) ] (\mathbf{n})
      > \epsilon_\text{supp},
      \,\, \mathbf{n} \in \mathbb{Z}^3 \right\}, \label{eq:d_supp} \\
    D_\text{soln} &= \left\{ B(\mathbf{n}) : [ W_\text{soln}(D_\infty) ] (\mathbf{n})
      > \epsilon_\text{soln},
      \,\, \mathbf{n} \in \mathbb{Z}^3 \right\}, \label{eq:d_soln}
  \end{align}
  \label{eq:d_defs}%
\end{subequations}
respectively.
The functions $W_\text{supp}$ and $W_\text{soln}$, and the scalars $\epsilon_\text{supp}$ and $\epsilon_\text{soln}$ are referred to as weight functions and threshold levels, respectively.

Although the weight function $W_\text{supp}$ can be defined to reflect any block selection criteria, we limit our attention to cases for which $[ W_\text{supp}(D_\infty) ] (\mathbf{n})$ reflects the magnitude of the fields $\mathsf{C}\mathsf{u}$ and $\mathsf{G}^\dagger \tilde{\mathsf{N}}(\mathsf{u}+\mathsf{u}_\infty)$ over the block $B(\mathbf{n})$.
This choice of $W_\text{supp}$ facilitates establishing relationships between the threshold level $\epsilon_\text{supp}$ and the error incurred by neglecting source terms values outside $D_\text{supp}$ when solving the discrete Poisson problems of Eq.\eqref{eq:proj_final} and \eqref{eq:vor_stf}.
As a representative example, we consider the weight function $W_\text{supp}$ given by
\begin{subequations}
  \begin{equation}
    [ W_\text{supp}(D_\infty) ] (\mathbf{n}) = \max\left(
      {\mu(\mathbf{n})}/{\mu_\text{global}},\,
      {\nu(\mathbf{n})}/{\nu_\text{global}} \right),
    \label{eq:weight_fun}
  \end{equation}
  \begin{alignat}{2}
    &\mu(\mathbf{n}) = \
      \max_{\mathbf{m} \in \text{\textsl{ind}}[B(\mathbf{n})]}
        | [\mathsf{C}\mathsf{u}](\mathbf{n})
          |, \quad
    &&\mu_\text{global} = \max_{\mathbf{n}\in\mathbb{Z}^3} \mu(\mathbf{n}) , \\
    &\nu(\mathbf{n}) = \
      \max_{\mathbf{m} \in \text{\textsl{ind}}[B(\mathbf{n})]}
        | [ \mathsf{G}^\dagger \tilde{\mathsf{N}}(\mathsf{u}+\mathsf{u}_\infty) ] (\mathbf{n})
          |, \quad
    &&\nu_\text{global} = \max_{\mathbf{n}\in\mathbb{Z}^3} \nu(\mathbf{n}).
  \end{alignat}%
\end{subequations}
In the absence of any error associated with computing the action of $\mathsf{L}_\mathcal{Q}^{-1}$, this expression for $W_\text{supp}$ results in an upper bound of $\epsilon_\text{supp}$ for the point-wise normalized residual of the active domain approximations of Eq.~\eqref{eq:proj_final} and \eqref{eq:vor_stf}.%
\footnote{%
Formally, $\epsilon_\text{supp}$ is only an approximate upper bound for the active domain case of Eq.~\eqref{eq:proj_final} since the source field for this problem is not exactly equal to $-\mathsf{G}^\dagger \tilde{\mathsf{N}}(\mathsf{u}+\mathsf{u}_\infty)$.
Yet, for the present error estimates, numerical experiments of representative flows indicate that $-\mathsf{G}^\dagger\tilde{\mathsf{N}}(\mathsf{u}+\mathsf{u}_\infty)$ at $t=t_k$ is a good approximation to $\mathsf{G}^\dagger \mathsf{r}_k^{i}$ of each stage of the $k$-th time-step.
}
For these cases, the point-wise normalized residual is defined as $\|\mathsf{r}\|_\infty/\|\mathsf{x}\|_\infty$, where
\begin{equation}
  \mathsf{r} = \mathsf{x} - \mathsf{M}^{\text{supp}} \mathsf{L}_\mathsf{Q} \mathsf{y},
  \quad
  \mathsf{y} = \mathsf{M}^{\text{xsoln}}_\mathcal{Q} {\mathsf{L}_\mathsf{Q}}^{-1}
    \mathsf{M}^{\text{supp}}_\mathcal{Q} \mathsf{x},
\end{equation}
and $\mathsf{x}$ is the source field of the corresponding discrete Poisson problem.

In general, as the solution changes over time the domain $D_\text{supp}$, as defined by Eq.~\eqref{eq:d_defs} and Eq.~\eqref{eq:d_supp}, will also change.
Significant amounts of non-negligible source terms are prevented from being advected or diffused outside $D_\text{supp}$ by recomputing and, if necessary, reinitializing the active domain at the beginning of a time-step.
This operation is performed by first computing $\mathsf{w}\leftarrow\mathsf{C}\mathsf{u}$ and $\mathsf{q}\leftarrow -\mathsf{G}^\dagger \tilde{\mathsf{N}}(\mathsf{u}+\mathsf{u}_\infty)$ on $D_\text{xsoln}$.
Next, values of $\mathsf{w}$ and $\mathsf{q}$ of grid cells belonging to block in $D_\text{buffer}$ that have been significantly contaminated by finite boundary errors are zeroed.
Finally, $[ W_\text{supp}(D_\infty) ] (\mathbf{n})$ and $[ W_\text{soln}(D_\infty) ] (\mathbf{n})$ are computed using Eq.~\eqref{eq:d_supp} for all $\mathbf{n}\in\mathbb{Z}^3$ such that $B(\mathbf{n}) \in D_\text{xsoln}$ and are set to zero otherwise.

If either of the newly computed $D_\text{supp}$ or $D_\text{soln}$ differ from their respective previous values, then it is necessary to reinitialize the active grid and compute the discrete velocity perturbation, $\mathsf{u}$, over the new $D_\text{xsoln}$.
By construction, all non-negligible values of the discrete vorticity, $\mathsf{w}$, are contained in $D_\text{supp}$; therefore, $\mathsf{u}$ over $D_\text{xsoln}$ can be computed as
\begin{equation}
  \mathsf{a} \leftarrow -\mathsf{M}^{\text{xsoln}}_\mathcal{E}
    {\mathsf{L}_\mathsf{Q}}^{-1} \mathsf{M}^{\text{supp}}_\mathcal{E} \mathsf{w},\quad
  \mathsf{u} \leftarrow \mathsf{M}^{\text{xsoln}}_\mathcal{E}
    \mathsf{C}^\dagger \mathsf{M}^{\text{xsoln}}_\mathcal{E} \mathsf{a}.
  \label{eq:vel_refresh}
\end{equation}
Subsequently, we denote the procedure implied by Eq.~\eqref{eq:vel_refresh} as $\mathsf{u} \leftarrow \text{Vor2Vel}( \mathsf{w} )$.

We emphasize that the present algorithm is also compatible with other choices of weight functions.
Using weight functions that are well-suited for capturing the relevant flow physics of a particular application can potentially reduce the size of the active domain and the number of operations required to accurately simulate the flow.
For example, if we are primarily interested in capturing the local physics of a flow over a particular region centered at $\mathbf{x}_0$, then a weight function $|\mathbf{n}-\mathbf{x}_0|^{-\alpha}[ W(D_\infty) ] (\mathbf{n})$ with $\alpha>0$ and $W$ given by Eq.~\eqref{eq:d_supp} might be an appropriate choice.
Unless otherwise stated, subsequent discussions assume that $W_\text{supp}$ is defined by Eq.~\eqref{eq:weight_fun}.

\subsection{Velocity refresh}
\label{sec:truncation_refresh}

In this section we present a set of techniques for limiting the error introduced from truncating non-compact fields that decay algebraically, e.g. $\mathsf{u}$ and $\mathsf{d}$, when computing the action of local operators.
We limit the present discussion to issues that arise from evaluating expressions involving $\mathsf{E}^{\mathsf{L}}_\mathcal{Q}(\alpha)$ on the finite active domain since this operator has the largest stencil of all local operators involved in the IF-HERK and projection methods.

We recall that the action of $\mathsf{E}^{\mathsf{L}}_\mathcal{Q}(\alpha)$ on $\mathsf{q}\in\mathbb{R}^\mathcal{Q}$ is computed as $[ \mathsf{G}_{\mathsf{E}}(\alpha) * \mathsf{q}](\mathbf{n})$.
Formally, $\mathsf{G}_{\mathsf{E}}(\alpha)$ has an infinite support, but, as discussed in Section~\ref{sec:spatial_lgfs}, $[\mathsf{G}_{\mathsf{E}}(\alpha)](\mathbf{n})$ decays rapidly as $|\mathbf{n}| \rightarrow \infty$; therefore, it is possible to approximate $\mathsf{G}_{\mathsf{E}}(\alpha)$ to prescribed tolerance using a finite support.
Consequently, for a given $\alpha$, there exists some $n_{\mathsf{E}}\in\mathbb{Z}$ such that the field induced from an arbitrary source field can be computed at a distance $n_{\mathsf{E}} \Delta x$ from $\partial D_\text{xsoln}$ to a prescribed accuracy $\epsilon_{\mathsf{E}}$.
By choosing the parameter $N_b$, used to define $D_\text{xsoln}$ in Eq.~\eqref{eq:def_xsoln}, to be equal or greater than $\lceil n_{\mathsf{E}} / n^b \rceil$ it is possible to evaluate the action of $\mathsf{E}^{\mathsf{L}}_\mathcal{Q}(\alpha)$ on $D_\text{soln}$ to an accuracy $\epsilon_{\mathsf{E}}$.
As a result, the flow inside $D_\text{soln}$ remains an accurate approximation of the flow that would have been obtained using the entire unbounded grid.

As the solution is evolved using the IF-HERK method, the operator $\mathsf{E}^{\mathsf{L}}_\mathcal{Q}(\alpha)$ is repeatedly applied to various grid functions, causing the error associated with truncated non-compact source fields to progressively propagate into the interior of $D_\text{xsoln}$.
The action of $\prod_{i=1}^{n} \mathsf{M}^{\text{xsoln}}_\mathcal{Q} \mathsf{E}^{\mathsf{L}}_\mathcal{Q}(\alpha_i) \mathsf{M}^{\text{xsoln}}_\mathcal{Q}$ is well-approximated by $\mathsf{M}^{\text{xsoln}}_\mathcal{Q} \mathsf{E}^{\mathsf{L}}_\mathcal{Q}(\beta) \mathsf{M}^{\text{xsoln}}_\mathcal{Q}$, where $\beta=\sum_{i=1}^{n} \alpha_i$.
Given that the physical values of the nonlinear terms in the IF-HERK algorithm are approximately zero on $D_\text{buffer}$, the minimum buffer region required to integrate $\mathsf{u}$ over $q$ time-steps is determined by the support of $\mathsf{G}_{\mathsf{E}}(q\beta)$, where $\beta=\sum_{i=1}^{s} \frac{\Delta \tilde{c}_i \Delta t}{(\Delta x)^2\text{Re}} = \frac{\Delta t}{(\Delta x)^2\text{Re}}$.
A procedure for obtaining estimates for $n_{\mathsf{E}}$ from $q$ and $\beta$ is provided in \ref{app:iferror}.
This procedure is extended to obtain an upper bound, $q_\text{max}$, on the number of time-steps, $q$, before the error at prescribed distance $n_{\mathsf{E}} \Delta x$ away from $\partial D_\text{xsoln}$ exceeds a prescribed value of $\epsilon_{\mathsf{E}}$.
At its minimum, the depth of the buffer region is $n^b N_b \Delta x$; therefore, the present method takes $n_{\mathsf{E}}$ to be equal to $n^b N_b$.

Provided $q_\text{max}\ge1$, the solution is integrated over multiple time-steps before the error from truncating non-compact source field starts to significantly affect the accuracy of the solution on $D_\text{soln}$.%
\footnote{%
Combinations of $n^b$, $N_b$, and $\beta$ resulting in $q_\text{max}=0$ are not allow. For a given $\beta$, the value of $q_\text{max}=0$ can always be increased by using larger values of $n^b$ or $N_b$.
}
In order to maintain the prescribed accuracy, after $q_\text{max}$ time-steps the discrete velocity perturbation on $D_\text{xsoln}$ is recomputed or \emph{refreshed} from the discrete vorticity on $D_\text{supp}$ using the $\text{Vor2Vel}$ procedure.

\section{Algorithm summary}
\label{sec:algorithm}

The present method for solving the incompressible Navier-Stokes on formally unbounded Cartesian grids using a finite number of operations and storage, referred to as the NS-LGF method, is summarized in this section.
Implementation details are omitted since they are beyond the scope of the present work.
Instead, we refer the reader to the parallel implementation of the LGF-FMM \cite{liska2014}, which can be readily extended to accommodate the additional operations required by the NS-LGF method.

An outline of the steps performed by the NS-LGF algorithm at $k$-th time-step is as follows:
\begin{enumerate}
  \item \emph{Preliminary}: compute the discrete vorticity, $\mathsf{w}_k$, and divergence of the Lamb vector, $\mathsf{q}_k$.
    \begin{subequations}
      \begin{align}
        \mathsf{w}_k &\leftarrow \mathsf{M}^\text{xsoln}_\mathcal{E} \mathsf{C} \mathsf{M}^\text{xsoln}_\mathcal{F} \mathsf{u}_k, \\
        \mathsf{q}_k &\leftarrow -\mathsf{M}^\text{xsoln}_\mathcal{C}
          \mathsf{G}^\dagger \mathsf{M}^\text{xsoln}_\mathcal{F}
            \tilde{\mathsf{N}}(\mathsf{M}^\text{xsoln}_\mathcal{F}(\mathsf{u}_k+\mathsf{u}_\infty(t_k))).
      \end{align}
    \end{subequations}
  \item \emph{Grid update}: update the computational grid based on prescribed criteria.
    \begin{enumerate}
      \item \emph{Query}: use weight functions $W_\text{supp}$ and $W_\text{soln}$, threshold values $\epsilon_\text{supp}$ and $\epsilon_\text{soln}$, and fields $\mathsf{w}_k$ and $\mathsf{q}_k$ to determine whether $D_\text{supp}$ or $D_\text{soln}$ need to be updated.
      \item \emph{Update}: (if necessary) update $D_\text{supp}$, $D_\text{soln}$, and $D_\text{xsoln}$ by adding or removing blocks.
      Copy the values of the discrete vorticity from the old to the new computational grid for $\forall B \in D^{\text{new}}_\text{supp} \cap D^{\text{old}}_\text{supp}$, where $D^{\text{new}}_\text{supp}$ and $D^{\text{old}}_\text{supp}$ denote $D_\text{supp}$ before and after the update, respectively.
    \end{enumerate}
  \item \emph{Velocity refresh}: compute the discrete velocity perturbation, $\mathsf{u}_k$, from the discrete vorticity, $\mathsf{w}_k$.
    \begin{enumerate}
      \item \emph{Query}: this operation is required if either the grid has been updated or if the number of time-steps since the last refresh is equal or greater than $q_\text{max}$.
      \item \emph{Refresh}: (if necessary) compute $\mathsf{u}_k$ using:
        \begin{equation}
          \mathsf{u}_k \leftarrow \text{Vor2Vel}( \mathsf{w}_k ),
        \end{equation}
      where the $\text{Vor2Vel}$ procedure given by Eq.~\eqref{eq:vel_refresh}.
    \end{enumerate}
  \item \emph{Time integration}: compute $\mathsf{u}_{k+1}$, $t_{k+1}$, and $\mathsf{p}_{k+1}$ using:
    \begin{equation}
      (\mathsf{u}_{k+1},t_{k+1},\mathsf{p}_{k+1})
        \leftarrow \text{xIF-HERK}(\mathsf{u}_{k},t_k),
    \end{equation}
    where the $\text{xIF-HERK}$ algorithm is the finite computational grid version of the $\text{IF-HERK}$ algorithm.
\end{enumerate}

The $\text{xIF-HERK}$ algorithm is identical to the $\text{IF-HERK}$ algorithm, except for the presence of mask operators which are used to confine all operations to the finite active domain.
With the exception of a few special cases, the $\text{xIF-HERK}$ algorithm is obtained by operating from the left all operators and grid functions present in the $\text{IF-HERK}$ algorithm by the appropriate $\mathsf{M}^\text{xsoln}_\mathcal{Q}$, e.g. $\mathsf{A} \rightarrow \mathsf{M}^\text{xsoln}_\mathcal{Q} \mathsf{A}$ and $\mathsf{y} \rightarrow \mathsf{M}^\text{xsoln}_\mathcal{Q} \mathsf{y}$.
The exceptions to this rule correspond to the expressions for $\mathsf{g}_k^i$ and $\hat{\mathsf{d}}_k^i$, which are given by
\begin{subequations}
  \begin{equation}
    \mathsf{g}_k^i = \tilde{a}_{i,i} \Delta t
      \mathsf{M}^\text{soln}_\mathcal{F} \tilde{\mathsf{N}}
        \left( \mathsf{M}^\text{xsoln}_\mathcal{F} ( \mathsf{u}_k^{i-1}
          + \mathsf{u}_\infty(t_k^{i-1}) ) \right),
    \label{eq:rhs_final}
  \end{equation}
  \begin{equation}
    \hat{\mathsf{d}}_k^{i} = - \mathsf{M}^\text{xsoln}_\mathcal{C}
      \mathsf{L}_\mathcal{C}^{-1} \mathsf{M}^\text{supp}_\mathcal{C}
        \mathsf{G}^\dagger \mathsf{M}^\text{xsoln}_\mathcal{F} \mathsf{r}_k^{i}.
    \label{eq:dpoisson_final}
  \end{equation}
\end{subequations}
Both Eq.~\eqref{eq:rhs_final} and \eqref{eq:dpoisson_final} reflect the fact that, by construction, the non-negligible physical values of $\mathsf{w}_k$ and $\mathsf{q}_k$ are contained in $W_\text{supp}$.

The operation count for the $k$-th time-step of the NS-LGF method, denoted by $N^{\text{NS}}_k$, is dominated by the number of operations required to evaluate the actions of $\mathsf{L}^{-1}_\mathsf{Q}$ and $\mathsf{E}^{\mathsf{L}}_\mathcal{Q}$.
As a result, an estimate for $N^{\text{NS}}_k$ is given by:
\begin{equation}
  N^{\text{NS}}_k
    \approx s N^{\mathsf{L}}_{k}
    + 3 C(s) N^{\mathsf{E}}_{k}
    + \lceil 3 N^{\mathsf{L}}_{k} \rfloor_k,
    \label{eq:op_count}
\end{equation}
where $s$ is the number of stages of the HERK scheme.
$N^{\mathsf{L}}_{k}$ and $N^{\mathsf{E}}_{k}$ denote the number of operations required to compute the action of $\mathsf{M}^\text{xsoln}_\mathcal{Q} \mathsf{L}^{-1}_\mathcal{Q} \mathsf{M}^\text{supp}_\mathcal{Q}$ and $\mathsf{M}^\text{xsoln}_\mathcal{Q} \mathsf{L}^{-1}_\mathcal{Q} \mathsf{M}^\text{xsoln}_\mathcal{Q}$, respectively, using the LGF-FMM for scalar grid spaces.%
\footnote{%
The factor of 3 that appears in the second and third terms of Eq.~\eqref{eq:op_count} accounts for the additional operations required to solve vector Poisson problems and vector integrating factors.
}
Detailed estimates for the values of $N^{\mathsf{L}}_{k}$ and $N^{\mathsf{E}}_{k}$ can be obtained from the discussion of the LGF-FMM \cite{liska2014}, but we note here that both $N^{\mathsf{L}}_{k}$ and $N^{\mathsf{E}}_{k}$ scale as $\mathcal{O}(N)$ for sufficiently large values of $N$, where $N$ is the total number of grid cells of the active domain.
The notation $\lceil\,\cdot\,\rfloor_k$ is used to clarify that cost associated with velocity update, i.e. $3 N^{\mathsf{L}}_{k}$, should only be included if a velocity update is performed.
Lastly, $C(s)$ specifies the number of integrating factors required by an $s$-stage $\text{IF-HERK}$ scheme.
In general, $C(s)$ is equal to $C_0(s)$, where
\begin{equation}
  C_0(s) = s + \left[ \frac{ (s-1) s }{ 2 } \right].
\end{equation}
For special case of second-order IF-HERK schemes, $C(s)$ reduces to $C_0(s)-1$.%
\footnote{%
The expression $c_s=1$ is one of the HERK order-conditions associated with second-order accurate constraints.
For the case of $c_s=\tilde{c}_{s-1}=1$, the integrating factor $\mathsf{H}_\mathcal{F}^s$, defined by Eq.~\eqref{eq:ifherk_aux_1}, simplifies to the identity operator.
}

For convenience, a summary of the parameters used in our treatment of the active computational domain is provided by Table~\ref{tab:run_parameters}.
\begin{table}[htbp]
  \centering
  \caption{%
  Parameters used in the treatment of the finite computational domain.
  \label{tab:run_parameters}
  }
  \begin{tabular}{|ccc|}
    \hline
    \emph{Symbol} & \emph{Description} & \emph{Section} \\ \hline
    $N_b$ & Width of $W_\text{buffer}$ (no. blocks) & \ref{sec:truncation_active} \\
    $n^{b}$ & Block size (no. cells) & \ref{sec:truncation_active} \\
    $\epsilon_\text{FMM}$ & LGF-FMM tolerance & \ref{sec:spatial_lgfs} \\
    $\epsilon_\text{supp}$ & Support region threshold & \ref{sec:truncation_adaptivity} \\
    $\epsilon_\mathsf{E}$ & Buffer region tolerance & \ref{sec:truncation_refresh} \\ \hline
  \end{tabular}
\end{table}
Of the parameters listed in Table~\ref{tab:run_parameters}, only $\epsilon_\text{FMM}$, $\epsilon_\mathsf{E}$, and $\epsilon_\text{supp}$ affect the accuracy of the numerical simulation.
The \emph{solution error} of the NS-LGF method, i.e. the error associated with approximately solving the fully discretized unbounded grid equations, is approximately bounded above by the sum of these three parameters.

The field values used to compute $D_\text{supp}$ should represent field values that would be obtained using the unbounded grid in the absence of numerical errors associated with the evaluation of discrete operators.
Spurious and unnecessary changes to the active domain are avoided by requiring
\begin{equation}
  \max( \epsilon_\text{FMM}, \epsilon_\mathsf{E} ) < \alpha \epsilon_\text{supp},
  \label{eq:cond_dgrowth}
\end{equation}
where $\alpha<1$ is a safety parameter specifying the sensitivity of the adaptive scheme to the solution errors associated with $\epsilon_\text{FMM}$ and $\epsilon_\mathsf{E}$.%
\footnote{%
Numerical experiments of representative flows have shown that $\alpha \approx 0.1$ is sufficiently small as to avoid most spurious and unnecessary changes to the computational grid.
}
Furthermore, using parameters that satisfy Eq.~\eqref{eq:cond_dgrowth} eliminates the inclusion of blocks that only contain field values that are on the same order as the solution error.

The values for $n^b$ and $N_b$ can also significantly affect the number of numerical operations performed by the NS-LGF method.
Smaller values of $n^b$ typically result in smaller active domains, but require more frequent velocity updates and often require the use of LGF-FMM schemes with less than optimal computational rates.
In practice, computationally efficient schemes are obtained by setting $N_b=1$ and determining the lower bound for $n^b$, denoted by $n_0^b$, from the prescribed value of $\epsilon_\mathsf{E}$.
Next, starting from $n_0^b$, progressively larger values of $n^b$ are considered until an efficient LGF-FMM scheme that achieves the prescribed $\epsilon_\text{FMM}$ tolerance is obtained.
The construction and computational performance of LGF-FMM schemes are discussed in \cite{liska2014}.

\section{Verification examples}
\label{sec:verif}

The behavior of the NS-LGF method is verified through numerical simulations of thin vortex rings.
We consider vortex rings of ring-radius $R$ and core-radius $\delta$, with circulation $\Gamma$ and Reynolds number $\text{Re}=\frac{\Gamma}{\nu}$, where $\nu$ is the kinematic viscosity of the fluid.
Unless otherwise stated, simulations are initiated with a vorticity distribution given by
\begin{equation}
  \omega_\theta(r,z) = \frac{\Gamma}{\pi \delta^2}
    \exp\left( \frac{ z^2 + (r-R)^2 } { \delta^2 } \right),
    \quad
  \omega_z(r,z) = 0,
  \label{eq:initial_omega_gauss}
\end{equation}
where $r=x^2+y^2$ and $\theta=\tan^{-1}(y/x)$.
As a result, the vortex ring initially translates in the positive $z$-direction due to its self-induced velocity \cite{saffman1992}.

The numerical experiments discussed in this section are initialized by first specifying an initial discrete vorticity, $\mathsf{w}_0$, and then using Eq.~\eqref{eq:vel_refresh} to obtain an initial discrete velocity perturbation, $\mathsf{u}_0$.
This procedure naturally leads to a $\mathsf{u}_0$ that is compatible with the IF-HERK method, i.e. $\mathsf{G}^\dagger \mathsf{u}_0=0$.
The initial active domain is chosen such that the $|\boldsymbol{\omega}|<10^{-10}$ outside the $D_\text{supp}$.
In order to avoid significant numerical artifacts due to the jump in the direction of the vorticity field at the ring origin, we limit our attention to vortex rings for which $|\boldsymbol{\omega}_\text{center}|<10^{-10}\max|\boldsymbol{\omega}|$, where $\boldsymbol{\omega}_\text{center}$ is the value of $\boldsymbol{\omega}$ at the center of the ring.
For the case of Eq.~\eqref{eq:initial_omega_gauss}, this condition is satisfied for $\delta/R<0.2$.

Provided a sufficiently large initial active domain, any sufficiently accurate process for computing $\mathsf{w}_0$ from $\boldsymbol{\omega}_0$ can be used to initialize the numerical simulations.
Yet it is convenient to use a process that naturally leads to a $\mathsf{w}_0$ such that $\mathsf{D}\mathsf{w}_0 \approx 0$.
In the absence of any numerical errors, $\tilde{\mathsf{w}}_0 = \mathsf{C}\mathsf{u}_0$ is equal to $\mathsf{w}_0$ if and only if $\mathsf{D}\mathsf{w}_0=0$.
For the case of $\mathsf{D}\mathsf{w}_0 \ne 0$, the support of $\tilde{\mathsf{w}}_0$ is typically larger than the support of $\mathsf{w}_0$, which in turn leads to larger active domains and complicates initial error estimates, i.e. $|\mathsf{w}_0|<\epsilon$ in $D_\text{supp}$ does not imply $|\tilde{\mathsf{w}}_0|<\epsilon$ in $D_\text{supp}$.
Provided $\nabla \cdot \boldsymbol{\omega}=0$, it is possible to construct $\mathsf{w}_0$ such that the magnitude of $\mathsf{D}\mathsf{w}_0$ is less than a prescribed tolerance by computing approximate values of the vorticity flux over the faces of the dual grid and applying the Divergence theorem to each dual cell.%
\footnote{%
The dual grid corresponds to a copy of the original staggered grid that has been shifted by half a grid cell in each direction. Cells, faces, edges, and vertices of the original grid can be regarded as vertices, edges, faces, and cells, respectively, of the dual grid.
}
For all test cases, a high-order quadrature scheme is used to integrate the initial vorticity distribution over the faces of the dual grid such that the resulting $\mathsf{w}_0$ satisfies $\|\mathsf{D}\mathsf{w}_0\|_\infty \approx 10^{-10}$.

Test cases are performed using $n^b=16$ and $N_b = 1$.
This choice of parameters leads to $\epsilon_\text{FMM}<10^{-8}$ for all values of $\Delta x$, $\Delta t$, and $\text{Re}$ considered.
The values of $\epsilon_\text{supp}$ and $\epsilon_\mathsf{E}$ are taken to be $\epsilon_\text{supp} = 0.1 \epsilon^*$ and $\epsilon_\mathsf{E} = \epsilon^*$.
The value of $\epsilon^*$ is varied across different sets of simulations, but is always such that $10^{-8} \le \epsilon^* \le 10^{-2}$.
The support domain $D_\text{supp}$ is computed using Eq.~\eqref{eq:d_supp} and Eq.~\eqref{eq:weight_fun}, and the solution domain $D_\text{soln}$ is set to be equal to $D_\text{supp}$.
It follows from our choice of parameters that the overall solution error is always bounded above by $\epsilon^*$.%
\footnote{%
The \emph{solution error}, as defined in Section~\ref{sec:algorithm}, should not be confused with the error of the solution.
}

With the exception of a few test cases discussed in Section~\ref{sec:verif-conv}, all numerical experiments are performed using the IF-HERK scheme denoted as ``Scheme A'' in Section~\ref{sec:temporal_ifherk}.
The time-step size, $\Delta t$, is held fixed during each simulation and chosen such that the $\text{CFL}$, based on the maximum point-wise velocity magnitude, does not exceed 0.75.
Unless otherwise stated, the freestream velocity, $\mathsf{u}_\infty$, is set to be zero.

\subsection{Discretization error}
\label{sec:verif-conv}

The order of accuracy of the discretization techniques is verified using spatial and temporal refinement studies on the early evolution of a vortex ring at $\text{Re}_0=\numprint{1000}$ with initial vorticity distributions given by
\begin{equation}
  \omega_\theta(r,z) = \left\{ \begin{array}{cl}
    \alpha \frac{\Gamma}{R^2} \exp\left( -4s^2/(R^2-s^2) \right) & \text{if}\,\,s \le R \\
    0 & \text{otherwise}
    \end{array} \right. , \quad
  \omega_z(r,z) = 0,
  \label{eq:initial_omega_bump}
\end{equation}
where $s^2 = z^2 + (r-R)^2$ and $\alpha$ is chosen such that $\omega_\theta$ integrates to $\Gamma$, i.e. $\alpha \simeq 0.54857674$.%
\footnote{%
The computational cost of the spatial convergence tests are reduced by using ``fat'' vortex rings such as those given by Eq.~\eqref{eq:initial_omega_bump}, which, unlike similar ``fat'' rings given by Eq.~\eqref{eq:initial_omega_gauss}, are continuous and differentiable at the origin.
}
Test cases are performed using fixed grids that are sufficiently large such that at any time-step of the simulation the active domain corresponds to a value of $\epsilon^*$ less than $10^{-8}$.

We use $\varepsilon_\mathbf{u} = \| \mathsf{u} - \mathsf{T}_\mathcal{F} \mathsf{u}^* \|_\infty / \| \mathsf{u}^* \|_\infty$ and $\varepsilon_\mathbf{p} = \| \mathsf{p} - \mathsf{T}_\mathcal{C} \mathsf{p}^* \|_\infty / \| \mathsf{p}^* \|_\infty$ to approximate the error at time $T$ of the velocity field, $\mathsf{u}$, and the pressure field, $\mathsf{p}$, respectively.
The superscript ${}^*$ is used to denote grid functions obtained from the test case with the highest resolution, i.e. smallest $\Delta x$ or $\Delta t$, included in the corresponding refinement study.
Point-wise comparisons between grid functions at different refinement levels are made possible through the use of the coarsening operators $\mathsf{T}_\mathcal{F}$ and $\mathsf{T}_\mathcal{C}$.
Finally, we define $\|\mathsf{x}\|_\infty$ as the maximum value of $|\mathsf{x}(\mathbf{n})|$ for all $\mathbf{n}$ associated with grid locations in $D_\text{soln}$.

\begin{figure}[htbp]
  \begin{center}
    \includegraphics[width=\textwidth]{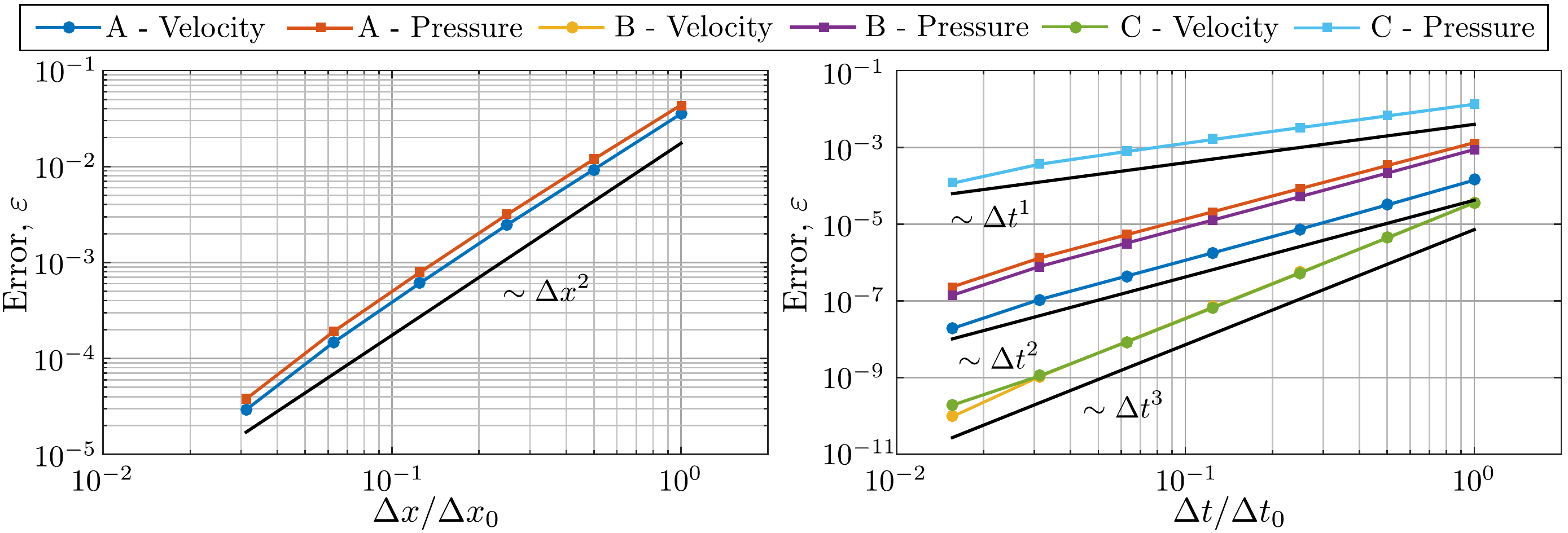}
  \end{center}
  \caption{%
  Velocity error, $\varepsilon_\mathbf{u}$, and pressure error, $\varepsilon_p$, for test cases. Spatial refinement study verifies second-order accuracy of the spatial discretization technique (\emph{left}).
  Temporal refinement studies verify the expected order of accuracy of the three time integration schemes defined in Section~\ref{sec:temporal_ifherk} (\emph{right}).
  \label{fig:verif-conv-xt}
  }
\end{figure}

The spatial refinement study consists of seven test cases corresponding to $\Delta x / \Delta x_0 = 2^{0}, 2^{-1},\dots,2^{-6}$.
Test cases are performed using the same $\Delta t$, and $\varepsilon_\mathbf{u}$ and $\varepsilon_\mathbf{p}$ are evaluated at $T=10\Delta t$.
The computational grids are constructed such that the location of vertices of coarser grids always coincide with the location of vertices of finer grids.
This enables the coarsened solution fields $T_\mathcal{C}\mathsf{p}^*$ and $T_\mathcal{F}\mathsf{u}^*$ to be computed by recursively averaging the values of the 8 (4) fine grid cells (faces) occupying the same physical region as the corresponding coarse grid cell (face).
The slope of the error curves depicted in the left plot of Figure~\ref{fig:verif-conv-xt} verifies that the solutions are second-order accurate in $\Delta x$.

Temporal refinement studies are performed using the three IF-HERK schemes, Scheme A--C, included in Section~\ref{sec:temporal_ifherk}.
For each scheme, a series of eight test cases is performed using $\Delta t / \Delta t_0 = 2^{0}, 2^{-1},\dots,2^{-7}$.
All test cases employ the same computational grid, and $\varepsilon_\mathbf{u}$ and $\varepsilon_\mathbf{p}$ are evaluated at $T=10\Delta t_0$.
Consequently, $\mathsf{T}_\mathcal{F}$ and $\mathsf{T}_\mathcal{C}$ are taken to be identity operators.
The slopes of the error curves depicted in the right plot of Figure~\ref{fig:verif-conv-xt} verify that the accuracy with respect to $\Delta t$ of each scheme is the same as the order of accuracy expected from the IF-HERK order-conditions.%
\footnote{%
We note that the spatial discretization error associated with the computational grid is significantly larger than the temporal discretization error for some test cases.
This does not affect the present refinement studies since the spatial discretization error is the same for all test cases and our error estimates are computed as the difference of two numerical solutions.
}

\subsection{Quality metrics for thin vortex rings}
\label{sec:verif-conv-int}

In this section we consider the laminar evolution of a thin vortex ring at $\text{Re}_0=\numprint{7500}$ initiated with $\delta_0 / R_0 = 0.2$.
Six test cases for different values of $\Delta x$ and $\Delta t$ are performed.
The ratio $\Delta t / \Delta x = 0.5734 R_0/\Gamma_0$ is held constant across all test cases.
Unlike the numerical experiments of Section~\ref{sec:verif-conv}, the grid is allowed to freely adapt as the solution evolves.
For all test cases, $\epsilon^*$ is taken to be $10^{-6}$, which is significantly smaller than the discretization error inferred from the discussion of Section~\ref{sec:verif-conv}.

The evolution of isolated vortex rings is often characterized by the time-history of a few fundamental volume integrals.
Quantities considered in the following numerical experiments include the hydrodynamic impulse $\boldsymbol{\mathcal{I}}$, the kinetic energy $\mathcal{K}$, enstrophy $\mathcal{E}$, the helicity $\mathcal{J}$, the Saffman-centroid $\boldsymbol{\mathcal{X}}$, and the ring-velocity $\boldsymbol{\mathcal{U}}$.
Expressions for these quantities for unbounded fluid domains and exponentially decaying $\boldsymbol{\omega}$ fields are given by \cite{saffman1992}:
\begin{align}
  \begin{split}
    \boldsymbol{\mathcal{I}}(t) &= \frac{1}{2} \int_{\mathbb{R}^3}
      \mathbf{x} \times \boldsymbol{\omega}\, d\mathbf{x}, \\
    \mathcal{K}(t) &= \int_{\mathbb{R}^3} \mathbf{u} \cdot
      \left( \mathbf{x} \times \boldsymbol{\omega} \right) \, d\mathbf{x}, \\
    \mathcal{E}(t) &= \frac{1}{2} \int_{\mathbb{R}^3}
      \left| \boldsymbol{\omega} \right|^2 \, d\mathbf{x}, \\
  \end{split} \quad
  \begin{split}
    \mathcal{J}(t) &= \int_{\mathbb{R}^3} \mathbf{u} \cdot \boldsymbol{\omega} \, d\mathbf{x}, \\
    \boldsymbol{\mathcal{X}}(t) &= \frac{1}{2} \int_{\mathbb{R}^3}
      \frac{ \left( \mathbf{x} \times \boldsymbol{\omega} \right) \cdot  \boldsymbol{\mathcal{I}} }
        { |\boldsymbol{\mathcal{I}}|^2 } \mathbf{x}\, d\mathbf{x}
    - \int_0^t \mathbf{u}_\infty(t^\prime)\,dt^\prime \\
    \boldsymbol{\mathcal{U}}(t) &= \frac{d\boldsymbol{\mathcal{X}}}{dt}.
  \end{split}
  \label{eq:int_quantities}
\end{align}
The hydrodynamic impulse, $\boldsymbol{\mathcal{I}}$, is a conserved quantity in the absence of non-conservative forces \cite{saffman1992}.
As a result, $\boldsymbol{\mathcal{I}}$ provides a useful metric for assessing the accuracy and physical fidelity of numerical solutions.
The time rate of change of $\mathcal{K}$ is related to $\mathcal{E}$ by the relationship $\frac{d}{dt}\mathcal{K} = - 2 \nu \mathcal{E}$.
Differences in the time history of $\frac{d}{dt}\mathcal{K}$ between different numerical simulations of the same flow are commonly used to characterize the accuracy of solutions of unsteady flows \cite{stanaway1988,archer2008,cheng2015}.
In the absence of viscosity, the helicity, $\mathcal{J}$, is an invariant of the flow and provides a measure for the degree of linkage of the vortex lines of the flow \cite{moffatt1992}.
Although the present simulations consider viscous flows, differences in $\mathcal{J}$ between test cases of the same flow are used as part of our quality metrics.
Our definitions for the vortex ring centroid, $\boldsymbol{\mathcal{X}}$, and propagation velocity, $\boldsymbol{\mathcal{U}}$, are equivalent to those used by Saffman \cite{saffman1970,saffman1992}.
Although all the integrals of Eq.~\eqref{eq:int_quantities} are formally over $\mathbb{R}^3$, they can be accurately computed for solutions obtained by the NS-LGF method since the support of the integrands is approximately contained in $D_\text{soln}$.%
\footnote{%
Numerical solutions set the vorticity outside the computational to be zero.
As a result, the only error involved in evaluating the integrals of Eq.~\eqref{eq:int_quantities} is the error resulting from their discretization.
}

\begin{figure}[htbp]
  \begin{center}
    \includegraphics[width=\textwidth]{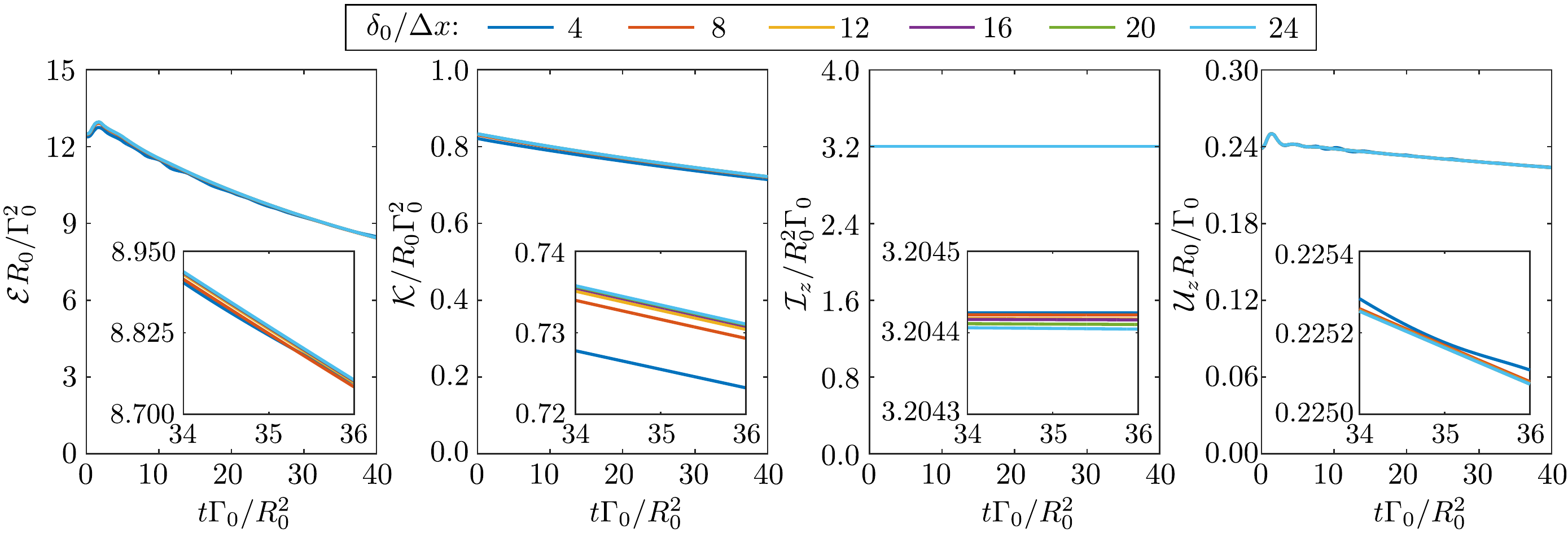}
  \end{center}
  \caption{%
  Time histories of $\mathcal{E}$, $\mathcal{K}$, $\mathcal{I}_z$, and $\mathcal{U}_z$ (\emph{respectively, left to right}) for a vortex ring at $\text{Re}_0=\numprint{7500}$ initiated with $\delta_0 / R_0 = 0.2$. Numerical experiments are performed using different values of $\delta_0 / \Delta x$ while holding $\Delta t / \Delta x$ constant.
  \label{fig:verif-conv-int}
  }
\end{figure}

\begin{table}[htbp]
  \centering
  \caption{%
  Maximum difference in $\mathcal{E}$, $\mathcal{K}$, $\mathcal{I}_z$, and $\mathcal{U}_z$ during $t \Gamma_0 / R_0^2 \in [ 0, 40 ]$ between test cases with $\delta_0 / \Delta x < 24$ and the test case with $\delta_0 / \Delta x = 24$.
  Reported differences have been normalized by the maximum value of the respective quantity during $t \Gamma_0 / R_0^2 \in [ 0, 40 ]$.
  \label{tab:verif-conv-int}
  }
  \begin{tabular}{c|cccc}
    $\delta_0 / \Delta x$ & $\mathcal{E}$ & $\mathcal{K}$ & $\mathcal{I}_z$ & $\mathcal{U}_z$ \\
    \hline
    $4$ & \
      $1.8\times10^{-2 }$ & \
      $1.5\times10^{-2 }$ & \
      $7.5\times10^{-6 }$ & \
      $4.9\times10^{-3 }$ \\
    $8$ & \
      $4.0\times10^{-3 }$ & \
      $3.5\times10^{-3 }$ & \
      $6.6\times10^{-6 }$ & \
      $4.8\times10^{-4 }$ \\
    $12$ & \
      $1.5\times10^{-3 }$ & \
      $1.3\times10^{-3 }$ & \
      $4.8\times10^{-6 }$ & \
      $1.7\times10^{-4 }$ \\
    $16$ & \
      $6.0\times10^{-4 }$ & \
      $5.3\times10^{-4 }$ & \
      $4.4\times10^{-6 }$ & \
      $7.3\times10^{-5 }$ \\
    $20$ & \
      $2.0\times10^{-4 }$ & \
      $2.2\times10^{-4 }$ & \
      $2.3\times10^{-6 }$ & \
      $2.7\times10^{-5 }$ \\
  \end{tabular}
\end{table}

The time history for the values of $\mathcal{E}$, $\mathcal{K}$, $\mathcal{I}_z$, and $\mathcal{U}_z$, where subscripts ``$q$'' denotes the component of a vector quantity in $q$-th direction, are shown in Figure~\ref{fig:verif-conv-int}.
The values for $\mathcal{J}$ and the components of $\boldsymbol{\mathcal{I}}$ and $\boldsymbol{\mathcal{U}}$ in the $x$- and $y$-directions were also computed, but are not depicted since the magnitude of these values remained less than $10^{-8}$, which is significantly smaller than $\epsilon^*$, for all test cases.
Visual inspection of the curves included in Figure~\ref{fig:verif-conv-int} suggests good agreement between all tests cases.
This is quantified by Table~\ref{tab:verif-conv-int}, which lists the maximum difference between test cases with $\delta_0 / \Delta x < 24$ and the test case with $\delta_0 / \Delta x = 24$.

Figure~\ref{fig:verif-conv-int} demonstrates that $\mathcal{E}$, $\mathcal{K}$, and $\mathcal{U}_z$ are most sensitive to changes in the resolution at early times, $t \Gamma_0 / R_0^2 \in [ 0, 15 ]$.
We attribute this to the rapid changes in the vorticity distribution observed shortly after the ring is initiated.
For cases initiated with finite values of $\delta/R$, it is well-known that flow undergoes an ``equilibration'' phase shortly after being initiated \cite{stanaway1988,shariff1994,archer2008}.%
\footnote{%
A vortex ring initiated a with vorticity distribution given by Eq.~\eqref{eq:initial_omega_gauss} is a solution to the Navier-Stokes equations only in the limit of $\delta/R \rightarrow 0$.%
}
During this phase, vorticity starts to be shed into the wake and, over time, the core region of the ring assumes a more relaxed axisymmetric vorticity distribution in which $\omega_\theta$ is no longer symmetric, but instead skewed so as to concentrate the vorticity away from the ring center.
After the equilibration phase, i.e. approximately after $t \Gamma_0 / R_0^2 > 15$ for test cases under consideration, the ring assumes a quasi-steady distribution that persists until the growth of linear instabilities causes the ring to transition into turbulence.
This transition does not occur during the simulation time of the present study, but will be investigated in Section~\ref{sec:verif-trunc}.

For each test case, the value of $\mathcal{I}$ remained nearly constant throughout the simulation time, only exhibiting deviations on the same order as $\epsilon^*$ (taken to be $10^{-6}$ for all test cases).
Interestingly, the value $\mathcal{I}$ appears to be insensitive to changes in $\Delta x$, at least when maintaining $\Delta t/\Delta x$ constant, as demonstrated by Table~\ref{tab:verif-conv-int}.
We refrain from speculating on whether the present method results in additional conservation properties beyond those mentioned in Section~\ref{sec:spatial_discrete}, since such investigations are beyond the scope of the present work.
Instead, we simply note that $\mathcal{I}$ appears to be conserved approximately up to the solution error, i.e. $\epsilon^*$, which further verifies the physical fidelity of solutions obtained using the NS-LGF method.

The difference between the LHS and RHS of $\frac{d}{dt}\mathcal{K} = -2 \nu \mathcal{E}$ is often used as a metric for the spatial discretization error.
The maximum value of $\left| \frac{d}{dt}\mathcal{K} - (-2 \nu \mathcal{E}) \right| / \left( 2 \nu \mathcal{E} \right)$ for $t \Gamma_0 / R_0^2 \in [ 0, 40 ]$ is $6.8 \times 10^{-2}$,  $2.1 \times 10^{-2}$, $9.6 \times 10^{-3}$, $5.3\times 10^{-3}$, $3.4\times 10^{-3}$, and $2.3\times 10^{-3}$ for the tests cases considered, sorted in ascending order of $\delta_0/\Delta x$.
Values for $\frac{d\mathcal{K}}{dt}$ and $2 \nu \mathcal{E}$ were computed at each half-time step using standard second-order differencing and averaging, respectively.

\subsection{Propagation speed of thin vortex rings}
\label{sec:verif-selfindc}

The results of this section verify that the solutions obtained using the NS-LGF method are indeed physical solutions to the incompressible Navier-Stokes equations.
The translational speed of laminar vortex rings has been extensively studied through experimental, numerical, and theoretical investigations \cite{saffman1992,stanaway1988,akhmetov2009,sullivan2008,fukumoto2010}.
\citet{saffman1970} showed that the propagation speed of a viscous vortex ring with a vorticity distributions given by Eq.~\eqref{eq:initial_omega_gauss}, in the limit of $\delta / R \rightarrow 0$, is
\begin{equation}
  U_\text{Saffman} = \frac{\Gamma_0}{4 \pi R_0} \left[
    \log \left( \frac{8}{\varepsilon} \right)
      - \beta_0
        + \mathcal{O} \left( \varepsilon \log \varepsilon \right) \right],
  \label{eq:u_saffman}
\end{equation}
where $\varepsilon = \delta / R$, $\beta_0 = \frac{1}{2} \left(1 - \gamma + \log 2 \right) \simeq 0.557966$, and $\gamma \simeq 0.577216$ is Euler's constant.
Subsequent numerical \cite{stanaway1988} and theoretical \cite{fukumoto2000} investigations have shown that the error term is actually smaller, and is given by $\mathcal{O} \left( \varepsilon^2 \log \varepsilon \right)$.

\begin{figure}[htbp]
  \begin{center}
    \includegraphics[width=\textwidth]{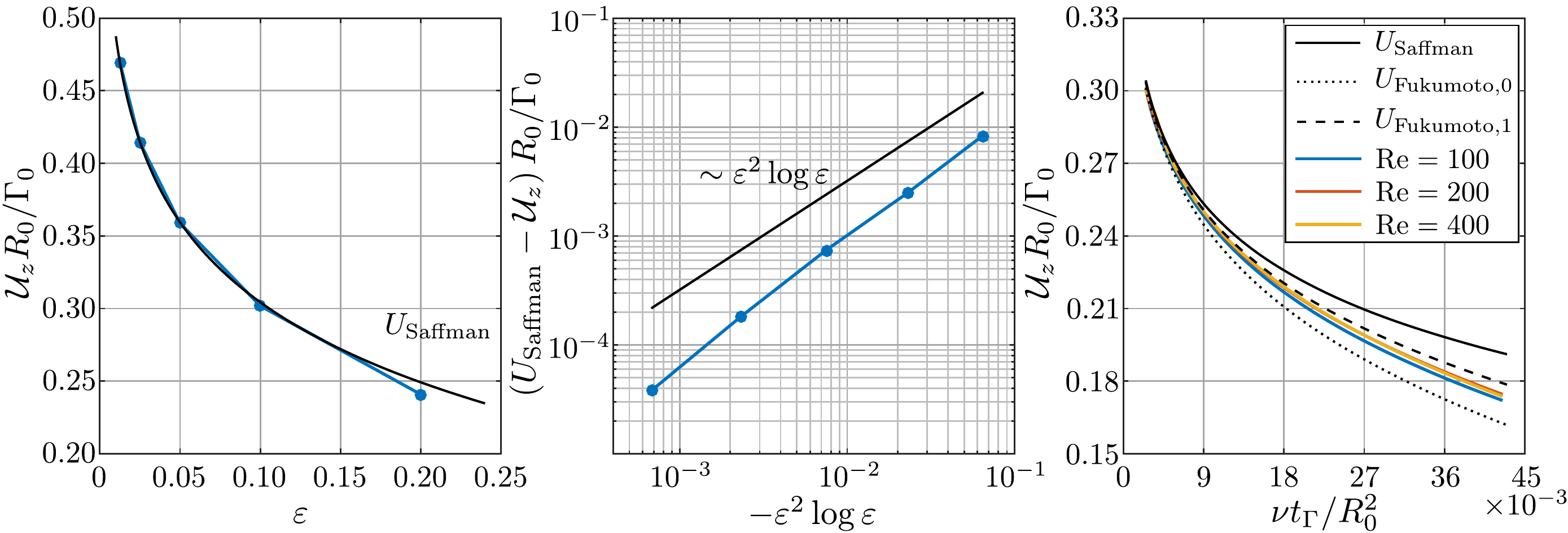}
  \end{center}
  \caption{%
  Propagation speed of a thin vortex ring at $\text{Re}_0=\numprint{7500}$ for the different values of $\varepsilon=\delta_0/R_0$ (\emph{left}).
  Difference between the computed value, $\mathcal{U}_z$, and the theoretical estimate, $U_\text{Saffman}$, for the propagation speed of a vortex ring at $\text{Re}_0=\numprint{7500}$ (\emph{middle}).
  Time history of the propagation speed of a vortex ring initiated with $\delta_0/R_0=0.1$ at different $\text{Re}$ (\emph{right}).
  \label{fig:verif-saff}
  }
\end{figure}

The initial propagation speed of a vortex ring, taken to be $\mathcal{U}_z$ as defined in Eq.~\eqref{eq:int_quantities}, is computed for test cases at $\text{Re}_0=\numprint{7500}$ that have been initiated with $\varepsilon = 0.2,\, 0.1,\, 0.05,\, 0.025$, and $0.0125$.
For all test cases, $\delta_0/{\Delta x}=20$, $\Delta t \Gamma_0 / R_0^2 = 10^{-6}$ and $\epsilon^*=10^{-6}$.
Values of $\mathcal{U}_z$ are computed via central differencing the values of $\boldsymbol{\mathcal{X}}$ between adjacent time-steps.
The value $\mathcal{U}_z$ at $t^* = \Delta t /2$ for each test case is shown in the left plot of Figure~\ref{fig:verif-saff}.
Visual inspection indicates good agreement between $\mathcal{U}_z$ and $U_\text{Saffman}$, which in turn verifies that numerical solutions obtained by the NS-LGF method approximate actual physical solutions.

We further verify the present formulation by confirming the form of the error term of $U_\text{Saffman}$, i.e. $\mathcal{O} \left( \varepsilon^2 \log \varepsilon \right)$.
Theoretical estimates for the effective ring and core radii for early times%
\footnote{%
The radius of the core and the vorticity centroid in the radial direction are approximately $2 \sqrt{vt}$ and $R_0 + 3 vt /R_0$ at $\sqrt{vt} \ll R_0$ \cite{fukumoto2010}.
}
indicate that, at time $t^*$, the ring and core size have not deviated enough from their initial values to significantly affect the value $U_\text{Saffman}$ as to hinder the present comparison.
The middle plot of Figure~\ref{fig:verif-saff} shows the difference in the ring propagation speed between the numerical experiments, $\mathcal{U}_z$, and theoretical estimates, $U_\text{Saffman}$.
For large values of $\varepsilon$, i.e. $\varepsilon>0.05$, the rate of change of $\Delta \tilde{U}_z = \left( U_\text{Saffman} - \mathcal{U}_z \right) R_0/\Gamma_0$ with respect to $\varepsilon$ is consistent with the theoretical $\mathcal{O} \left( \varepsilon^2 \log \varepsilon \right)$ error estimate.
On the other hand, for $\varepsilon<0.05$ the rate of change of $\Delta \tilde{U}_z$ with respect to $\varepsilon$ is slightly faster than $\mathcal{O} \left( \varepsilon^2 \log \varepsilon \right)$.
We refrain from attributing any physical meaning to the difference in the behavior of the error at smaller values of $\varepsilon$ since we have not thoroughly determined the numerical error for such test cases.%
\footnote{%
Extrapolating from the results of Table~\ref{tab:verif-conv-int} to the present tests cases, we estimate that the error of $\mathcal{U}_z$ to be between $10^{-5}$ and $10^{-4}$. As a result, the assumption that $\mathcal{U}_z$ is more accurate than $U_\text{Saffman}$ might need to be revisited for test cases resulting in values of $\Delta \tilde{U}_z<10^{-4}$.
}

We further verify the present implementation by comparing the time and Reynolds number dependence of $\mathcal{U}_z$ with previously reported theoretical \cite{fukumoto2010} and numerical \cite{stanaway1988} results.
To facilitate the comparisons, it is convenient to define
\begin{equation}
  t_\Gamma = \frac{\delta_0^2}{4\nu} + t.
\end{equation}
The discussion of \cite{fukumoto2010} provides theoretical bounds on $\mathcal{U}_z$ based on the low and high $\text{Re}$ limits of a vortex ring initiated with $\delta/R \rightarrow 0$,
\begin{subequations}
  \begin{alignat}{2}
    U_\text{Fukumoto,0} &= \frac{\Gamma_0}{4 \pi R_0} \left[
      \log \left( \frac{4R_0}{\sqrt{\nu t_\Gamma}} \right)
        - \beta_0 - \frac{9}{5}\left(
          \log \left( \frac{4R_0}{\sqrt{\nu t_\Gamma}} \right)
            -\beta_1 \right) \frac{\nu t_\Gamma}{R_0^2} \right]
   && (\text{low-Re}), \\
    U_\text{Fukumoto,1} &= \frac{\Gamma_0}{4 \pi R_0} \left[
      \log \left( \frac{4R_0}{\sqrt{\nu t_\Gamma}} \right)
        - \beta_0 - \beta_2 \frac{\nu t_\Gamma}{R_0^2} \right]
  && (\text{high-Re}),
  \end{alignat}
\end{subequations}
where $\beta_0$ is the same as in Eq.~\eqref{eq:u_saffman}, $\beta_1 \simeq 1.057967$, and $\beta_2 \simeq 3.671591$.
For all test cases, $\delta_0 / \Delta x= 15$ and $\Delta t$ is determined by requiring the initial CFL to be $0.5$.
Test cases correspond to a vortex ring at $\text{Re}_0 = 100,\, 200,\, \text{and}\,\, 400$ that are initiated with $\delta_0/R_0=0.1$.
The right plot of Figure~\ref{fig:verif-saff} demonstrates that, for all test cases, $\mathcal{U}_z$ remains bounded between $U_\text{Fukumoto,0}$ and $U_\text{Fukumoto,1}$, except at early times for the case of $\text{Re}_0=400$ where the numerical $\mathcal{U}_z$ slightly exceeds the $U_\text{Fukumoto,1}$.
This discrepancy is not surprising since the theory of \citet{fukumoto2010} assumes that the vortex ring is initiated with $\delta/R \rightarrow 0$, and, as a result, does not properly account for the changes in the vorticity distribution that occur during the equilibration phase of a vortex ring initiated with a finite $\delta/R$ value.
Although not shown in Figure~\ref{fig:verif-saff}, the time history of $\mathcal{U}_z$ for all test cases has been compared to the numerical results of \cite{stanaway1988}, and found to be in good agreement (overlaying the curves of both investigations reveal nearly identical results).

\subsection{Finite active computational domain error}
\label{sec:verif-trunc}

In this section, we investigate the effect that our adaptive grid technique has on the numerical solutions by considering the evolution of a thin vortex ring computed using different values of $\epsilon^*$.
These test cases are used to verify that the solutions converge as $\epsilon^*$ tends to zero and to verify, via comparisons with numerical investigations of other authors, the physical fidelity of the solutions.

For all test cases, the vortex ring is initiated with $\delta_0/R_0=0.2$ and a constant uniform flow, $\mathsf{u}_\infty = \left[ 0, 0, u^{(z)}_\infty \right]$, is superimposed to partially oppose the translational motion of the vortex ring.
The value of $u^{(z)}_\infty R_0 / \Gamma_0$ is taken to be $-0.18686$, which reduces the initial speed of the vortex ring by approximately 75\%.
Solutions are computed using $\delta_0/{\Delta x}=10$ and $\Delta t \Gamma_0 / R_0^2 \approx 0.01721$.
The error estimates of Section~\ref{sec:verif-conv-int} indicate that, for all test cases, the discretization error is on the order of $10^{-3}$.

\begin{figure}[htbp]
  \begin{center}
    \includegraphics[width=\textwidth]{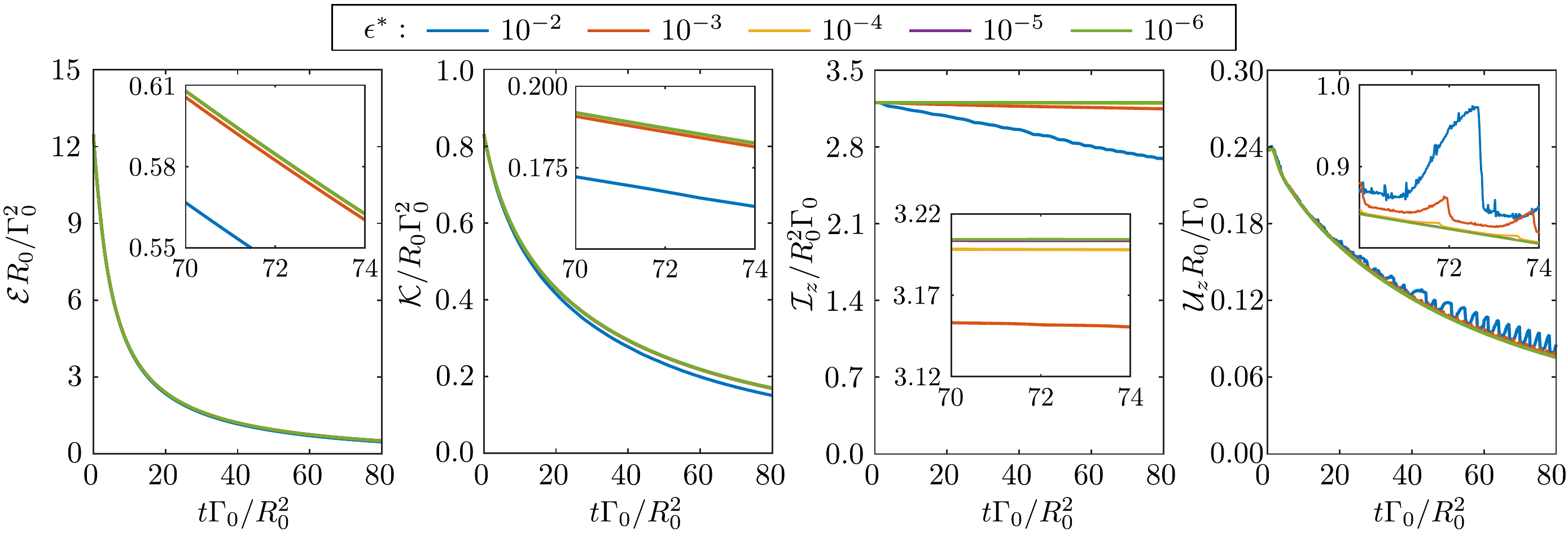}
  \end{center}
  \caption{%
  Time histories of $\mathcal{E}$, $\mathcal{K}$, $\mathcal{I}_z$, and $\mathcal{U}_z$ (\emph{respectively, left to right}) for a vortex ring at $\text{Re}_0=500$ initiated with $\delta_0 / R_0 = 0.2$. All parameters, with the exception of $\epsilon^*$, are held constant across all test cases.
  \label{fig:verif-conv-lowre}
  }
\end{figure}

Figure~\ref{fig:verif-conv-lowre} depicts the time histories of $\mathcal{E}$, $\mathcal{K}$, $\mathcal{I}_z$, and $\mathcal{U}_z$ for a vortex ring at $\text{Re}=500$ computed using $\epsilon^* = 10^{-2},\, 10^{-3},\, 10^{-4},\, 10^{-5},\, \text{and}\,\, 10^{-6}$.
The smooth decay of $\mathcal{E}$ and $\mathcal{K}$ indicates that the vortex ring remains laminar throughout the entire simulation time.
This follows from the fact that a pronounced peak in $\mathcal{E}$ is observed during the transition to the early stages of turbulence resulting from a significant increase in the stretching of vortex filaments \cite{archer2008}.
Figure~\ref{fig:verif-conv-lowre} verifies that, for laminar flows, numerical solutions converge as $\epsilon^*$ tends to zero.
For all test cases with values of $\epsilon^*>10^{-2}$, the error%
\footnote{
The error is estimated by assuming that the test case corresponding to $\epsilon^*=10^{-6}$ is the true solution.
}
in the computed values $\mathcal{E}$, $\mathcal{K}$ and $\mathcal{I}_z$ is inversely proportional to $\epsilon^*$ for $t \Gamma_0/R_0^2\in[10,80]$.
The large oscillations in $\mathcal{U}_z$ are due to shifts in $\boldsymbol{\mathcal{X}}$ resulting from the addition or removal of a single layer blocks in the $z$-direction.
For times at which all test cases exhibit an approximate local minimum in $\mathcal{U}_z$, e.g. $t \Gamma_0/R_0^2 \approx 70.5$, the error in $\mathcal{U}_z$ is also inversely proportional to $\epsilon^*$.

Next, we consider the effect $\epsilon^*$ has on solutions of unsteady flows that are sensitive to small perturbations.
The numerical investigations of \cite{bergdorf2007,archer2008} on thin vortex rings with Gaussian vorticity distributions at $\text{Re}_0=\numprint{7500}$ have shown that small sinusoidal perturbations to the vortex ring centerline result in the growth of azimuthal instabilities, which in turn facilitate the laminar to turbulent transition of the flow.
Here, we consider the evolution of a vortex ring at $\text{Re}_0=\numprint{7500}$ computed using values of $\epsilon^* = 10^{-2},\, 10^{-3},\, 10^{-4},\, 10^{-5},\, \text{and}\,\, 10^{-6}$.
Unlike the numerical experiments of \cite{bergdorf2007,archer2008}, the vortex ring is initiated without imposing any perturbations beyond those implied by the numerical scheme.

\begin{figure}[htbp]
  \begin{center}
    \includegraphics[width=\textwidth]{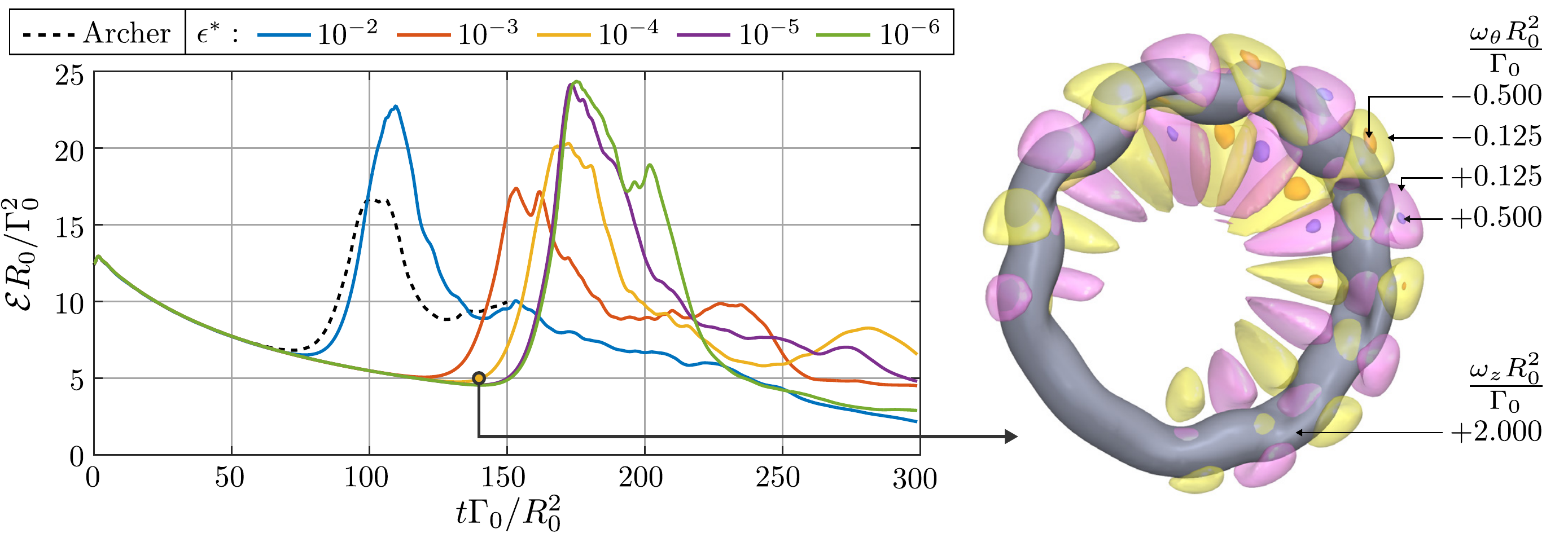}
  \end{center}
  \caption{%
  Time history of $\mathcal{E}$ for a vortex ring at $\text{Re}_0=\numprint{7500}$ initiated with $\delta_0 / R_0 = 0.2$ (\emph{left}). Data points labeled as ``Archer'' correspond values reported in \citet{archer2008}. All parameters, with the exception of $\epsilon^*$, are held constant across all test cases. Vorticity iso-surfaces at $t \Gamma_0/R_0^2=137.6$ for test case $\epsilon^*=10^{-4}$ (\emph{right}).
  \label{fig:verif-conv-thres}
  }
\end{figure}

The time history of $\mathcal{E}$ for all test cases is shown in the left plot of Figure~\ref{fig:verif-conv-thres}.
The transition into the early stages of turbulence, characterized by a peak in $\mathcal{E}$ resulting from an increase in the stretching of vortex filaments, is observed for all test cases.
The growth of azimuthal instabilities and the development of secondary or ``halo'' vortices occurring at beginning of the transition phase \cite{bergdorf2007,archer2008} are depicted in the right plot of Figure~\ref{fig:verif-conv-thres}.

As expected from the previous test cases for $\text{Re}_0=500$, the values of $\mathcal{E}$ during the laminar regime for all test cases converge as $\epsilon^*$ tends zero.
Also included in Figure~\ref{fig:verif-conv-thres} are the values of $\mathcal{E}$ reported in the numerical investigations of \citet{archer2008} for same vortex ring, which are nearly identical to values obtained from our test cases during the laminar regime.%
\footnote{%
In the discussion of \citet{archer2008}, the test case corresponding to a vortex ring at $\text{Re}_0=\numprint{7500}$ initiated with $\delta_0/R_0=0.2$ is denoted as case ``B3''.
Unlike the present test cases, the initial vorticity distribution for case B3 of \citet{archer2008} was slightly perturbed to promote an early transition.
}
Additionally, the vorticity iso-surfaces shown in right plot of Figure~\ref{fig:verif-conv-thres} are qualitatively similar to the vorticity iso-surfaces provided by \citet{archer2008} depicting the nonlinear growth of instabilities.
In particular, the iso-surfaces of both investigations demonstrate the noticeable presence of the $n=1$ azimuthal Fourier mode and the presence of halo vortices (iso-surfaces of $\omega_z$ in Figure~\ref{fig:verif-conv-thres}) of similar magnitudes but alternating sign wedged between the approximately sinusoidally displaced inner-core (iso-surfaces of $\omega_\theta$ in Figure~\ref{fig:verif-conv-thres}).

\begin{figure}[htbp]
  \begin{center}
    \includegraphics[width=\textwidth]{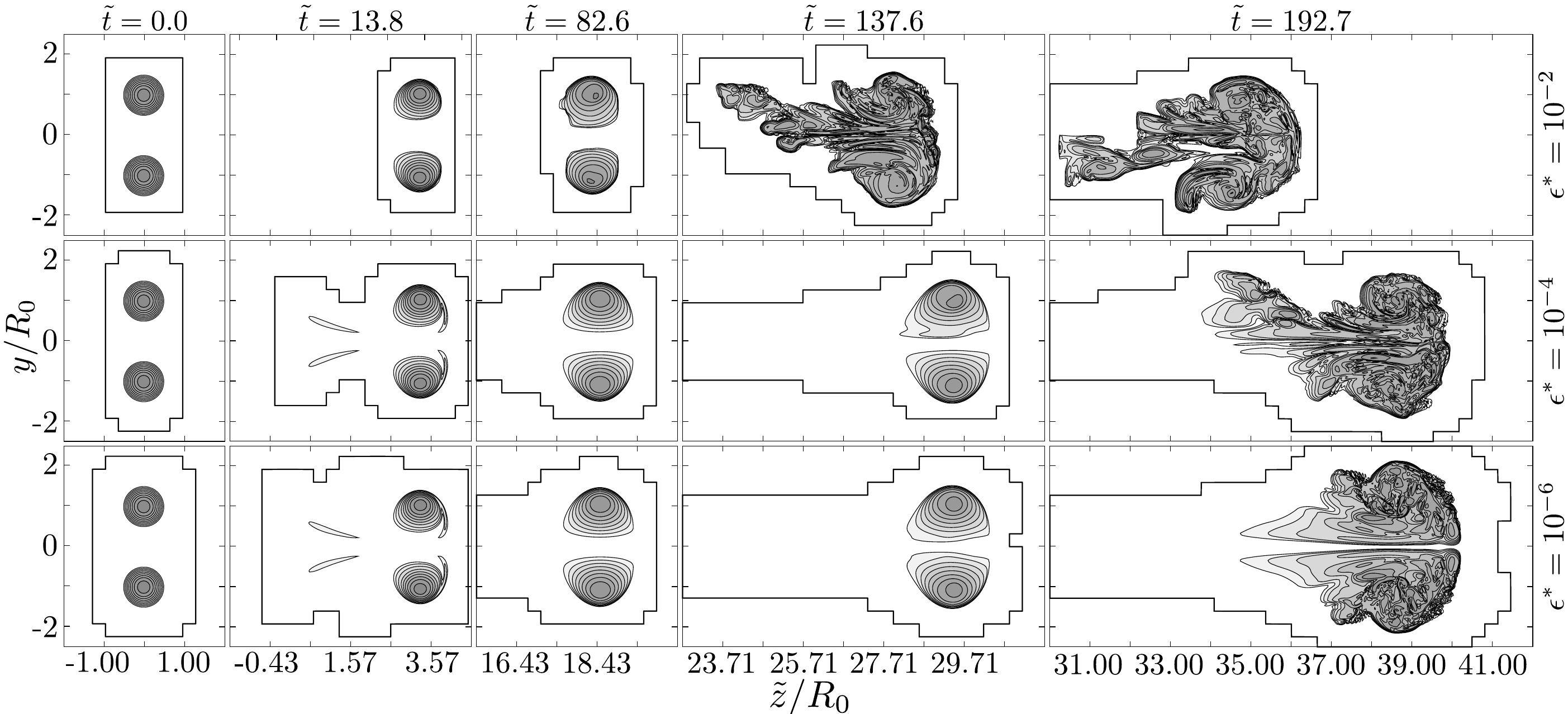}
  \end{center}
  \caption{%
  Vorticity magnitude on the $y$-$z$ plane at $x=0$ for test cases of $\epsilon^* = 10^{-2},\, 10^{-4},\, \text{and}\,\, 10^{-6}$ at different times, $\tilde{t}=t \Gamma_0/R_0^2$. Contours correspond to values of $|\omega|R_0^2/\Gamma_0 = 4 \times \left( \frac{1}{2} \right)^i$ for $i=8,7,\dots,0$. Contours have been shifted the $z$-direction to account for the constant freestream velocity, $\tilde{z} = z - u^{(z)}_\infty t$. Thick lines depict the boundary of $D_\text{xsoln}$.
  \label{fig:verif-contours}
  }
\end{figure}

The time histories of $\mathcal{E}$ shown in Figure~\ref{fig:verif-conv-thres} indicate that the time at which $\mathcal{E}$ starts to increase prior to reaching its peak value, i.e. the time at which the flow starts to transition, increases as $\epsilon^*$ decreases, but converges as $\epsilon^*$ tend to zero.
This trend is an expected consequence of the present adaptive grid technique since the flow field is slightly perturbed each time a block is removed, i.e. vorticity is implicitly set to zero outside $D_\text{supp}$.
The magnitude of these perturbations is correlated to the value of $\epsilon^*$ used to compute the numerical solution.
Over time, the perturbations introduced by the adaptive grid lead to changes in the flow field that break the axial symmetry of the solution, which in turn promotes the growth of instabilities.
Figure~\ref{fig:verif-contours} provides vorticity contours at different times that depict the breakdown of axial symmetry and the subsequent laminar to turbulent transition for a few test cases.

Figure~\ref{fig:verif-contours} also depicts the computational domains that result from using different values of $\epsilon^*$.
As expected, higher values of $\epsilon^*$ result in tighter domains, but lead to some significant changes in the flow that are potentially relevant to specific applications.
For example, Figure~\ref{fig:verif-contours} indicates that using a value $\epsilon^*$ of $10^{-2}$ is sufficient to accurately track the laminar evolution of the vortex core, but does not adequately capture the large wake that develops behind the vortex ring.%
\footnote{%
The maximum length, in terms of $R_0$, of the computational in the $z$-direction for is approximately $10$, $26$, $34$, $46$, $46$ for test case with $\epsilon^*$ equal to $10^{-2}$, $10^{-3}$, $10^{-4}$, $10^{-5}$, and $10^{-6}$, respectively.
}
We recall that the computational domain is determined by the particular choice of $W_\text{supp}$ and $\epsilon_\text{supp}$, both of which can be readily modified to accurately and efficiently capture the relevant physics of specific applications.

\begin{figure}[htbp]
  \begin{center}
    \includegraphics[width=\textwidth]{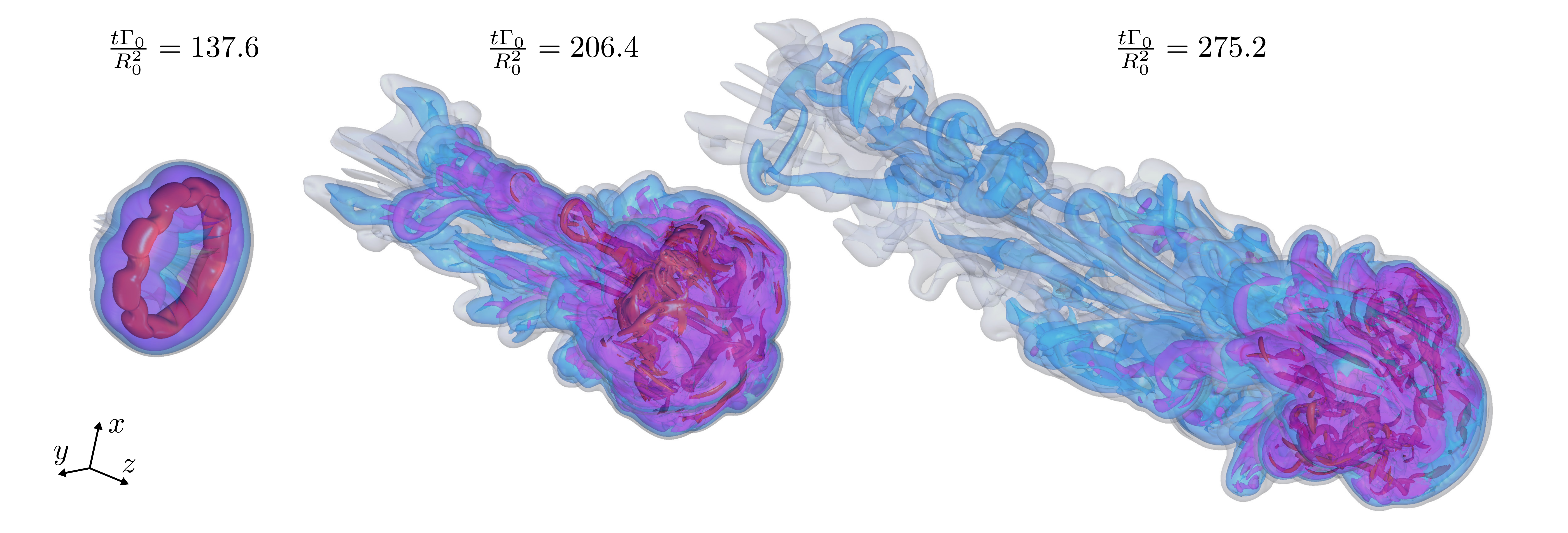}
  \end{center}
  \caption{%
  Translucent iso-surfaces of the vorticity magnitude for the test case of $\epsilon^*=10^{-4}$ at different times. Iso-surfaces correspond to values of $|\omega|R_0^2/\Gamma_0 = 0.03125,\, 0.125,\, 0.5,\, \text{and}\,\, 2$.
  \label{fig:verif-isos}
  }
\end{figure}

Figure~\ref{fig:verif-isos} depicts vorticity iso-surfaces during the transition phase ($t \Gamma_0 / R_0^2 = 137.6$) and early turbulent regime ($t \Gamma_0 / R_0^2 = 206.4\,\, \text{and}\,\, 275.2$) for the test case of $\epsilon^*=10^{-4}$.
At $t \Gamma_0 / R_0^2 = 206.4$ and $275.2$, the presence of multiple thin vortex filaments and the absence of a coherent core indicate that the vortex ring is in its early turbulent regime \cite{bergdorf2007,archer2008}.
A comparison of the vorticity iso-surfaces at $t \Gamma_0 / R_0^2 = 206.4$ and at $t \Gamma_0 / R_0^2 = 275.2$ demonstrates that interwoven vorticity filaments near the core region are gradually pushed into the wake.
As some of these structures are convected into the wake, they form hairpin vortices which persist for some time in the wake region.
The periodic shedding of hairpin vortices into the wake is consistent with the numerical investigations of \cite{bergdorf2007,archer2008}, which in turn further verifies the physical fidelity of our solutions.

\section{Conclusions}

We have reported on a new fast, parallel solver for 3D, viscous, incompressible flows on unbounded domains based on LGFs.
In this method, the incompressible Navier-Stokes equations are formally discretized on an unbounded staggered Cartesian grid using a second-order finite-volume scheme.
This discretization technique has the advantage of enforcing discrete conservation laws and producing discrete operators with mimetic and commutativity properties that facilitate the implementation of fast, robust solvers.
The system of DAEs resulting from the spatial-discretization of the momentum equation and the incompressibility constraint are integrated in time by using an integrating factor technique for the viscous terms and a HERK scheme for the convective term and the incompressibility constraint.
Computationally efficient expressions for the integrating factors are obtained via Fourier analysis on unbounded Cartesian grids.
A projection method that takes advantage of the mimetic and commutativity properties of the discrete operators is used to efficiently solve the linear system of equations arising at each stage of the time integration scheme.
This projection technique has the advantage of being equivalent to the LU decomposition of the system of equations, and, as a result, does not introduce any splitting-error and does not change the stability of the discretized equations.

In our formulation, solutions to the discrete Poisson problems and integration factor that are required to advance the flow are obtained through LGF techniques.
These techniques express the solutions to inhomogeneous difference equations as the discrete convolution between source terms and the fundamental solutions of the discrete operators on unbounded regular grids.
Fast, parallel solutions to the expressions resulting from the application of LGF techniques to discrete Poisson problems and integrating factors are obtained using the FMM for LGFs of \cite{liska2014}.

As a result of our LGF formulation, the flow is solved using only information contained in the grid region where the vorticity and the divergence of the Lamb vector have non-negligible values.
An adaptive block-structured grid and a velocity refresh technique are used to limit operations to a small finite computational domain.
In order to efficiently compute solutions to a prescribed tolerance, weight functions and threshold values are used to determine the behavior of the adaptive grid.

The order of accuracy of the discretization and solution techniques is verified through refinement studies.
The physical fidelity of the method is demonstrated in comparisons between computed and theoretical values for the propagation speed of a thin vortex ring.
Additionally, results for the evolution of a thin vortex ring at $\text{Re}_0=\numprint{7500}$ from the laminar to the early turbulent regime are shown to be in good agreement with investigations of other authors.

\section*{Acknowledgments}

This work was partially supported by the United States Air Force Office of Scientific Research (FA950--09--1--0189) and the Caltech Field Laboratory for Optimized Wind Energy with Prof. John Dabiri as PI under the support of the Gordon and Betty Moore Foundation.

\appendix
\section{Discrete operators}
\label{app:oprs}

In this appendix we provide point-operator and Fourier representations for the discrete operators of the present formulation.
For operators that map onto $\mathbb{R}^\mathcal{F}$ or $\mathbb{R}^\mathcal{E}$, expressions for only one component of the resulting vector fields are provided since expressions for the other components are readily deduced.
In the following discussion $\mathsf{c}\in\mathbb{R}^\mathcal{C}$, $\mathsf{f}\in\mathbb{R}^\mathcal{F}$, $\mathsf{e}\in\mathcal{E}$, and $\mathsf{v}\in\mathbb{R}^\mathcal{V}$ are arbitrary grid functions.

Point-operator representation based on the indexing convention depicted in Figure~\ref{fig:grid-cell} are as follows:

\begin{itemize}
\item\emph{Discrete gradient operators}: %
  $\mathsf{G} : \mathbb{R}^\mathcal{C} \mapsto \mathbb{R}^\mathcal{F}$ and %
  $\overline{\mathsf{G}}=-\mathsf{D}^\dagger : \mathbb{R}^\mathcal{V} \mapsto \mathbb{R}^\mathcal{E}$, where
\begin{subequations}
  \begin{align}
    \Delta x [ \mathsf{G} \mathsf{c} ]_{i,j,k}^{(1)}
      &= \mathsf{c}_{i+1,j,k} - \mathsf{c}_{i,j,k},
    \\
    \Delta x [ \overline{\mathsf{G}} \mathsf{v} ]_{i,j,k}^{(1)}
      &= \mathsf{v}_{i,j,k}^{(1)}
        - \mathsf{v}_{i-1,j,k}^{(1)}.
  \end{align}
\end{subequations}

\item\emph{Discrete curl operators}: $\mathsf{C} : \mathbb{R}^\mathcal{F} \mapsto \mathbb{R}^\mathcal{E}$ and %
  $\overline{\mathsf{C}}=\mathsf{C}^\dagger : \mathbb{R}^\mathcal{E} \mapsto \mathbb{R}^\mathcal{F}$, where
\begin{subequations}
  \begin{align}
    \Delta x [ \mathsf{C} \mathsf{f} ]_{i,j,k}^{(1)}
      &= \mathsf{f}_{i,j,k}^{(2)} - \mathsf{f}_{i,j,k+1}^{(2)}
        + \mathsf{f}_{i,j+1,k}^{(3)} - \mathsf{f}_{i,j,k}^{(3)},
    \\
    \Delta x [ \overline{\mathsf{C}} \mathsf{e} ]_{i,j,k}^{(1)}
      &= \mathsf{e}_{i,j,k-1}^{(2)} - \mathsf{e}_{i,j,k}^{(2)}
        + \mathsf{e}_{i,j,k}^{(3)} - \mathsf{e}_{i,j-1,k}^{(3)}.
  \end{align}
\end{subequations}

\item\emph{Discrete divergence operators}: $\mathsf{D} : \mathbb{R}^\mathcal{E} \mapsto \mathbb{R}^\mathcal{V}$ and %
  $\overline{\mathsf{D}}=-\mathsf{G}^\dagger : \mathbb{R}^\mathcal{F} \mapsto \mathbb{R}^\mathcal{C}$, where
\begin{subequations}
  \begin{align}
    \Delta x [ \mathsf{D} \mathsf{e} ]_{i,j,k}
      &= 
          \mathsf{e}_{i+1,j,k}^{(1)}
        + \mathsf{e}_{i,j+1,k}^{(2)}
        + \mathsf{e}_{i,j,k+1}^{(3)}
        - \sum_{q=1}^{3} \mathsf{e}_{i,j,k}^{(q)},
    \\
    \Delta x [ \overline{\mathsf{D}} \mathsf{f} ]_{i,j,k}
      &= 
        - \mathsf{f}_{i-1,j,k}^{(1)}
        - \mathsf{f}_{i,j-1,k}^{(2)}
        - \mathsf{f}_{i,j,k-1}^{(3)}
        + \sum_{q=1}^{3} \mathsf{f}_{i,j,k}^{(q)}.
  \end{align}
\end{subequations}

\item\emph{Discrete Laplace operators}: $\mathsf{L}_\mathcal{Q}:\mathbb{R}^\mathcal{Q} \mapsto \mathbb{R}^\mathcal{Q}$ %
		for all $\mathcal{Q}$ in $\{\mathcal{C},\mathcal{F},\mathcal{E},\mathcal{V}\}$, where
\begin{equation}
  \mathsf{L}_\mathcal{C} = -\mathsf{G}^\dagger \mathsf{G},
  \quad
  \mathsf{L}_\mathcal{V} = -\mathsf{D} \mathsf{D}^\dagger,
  \quad
  \mathsf{L}_\mathcal{F} = -\mathsf{G} \mathsf{G}^\dagger - \mathsf{C}^\dagger \mathsf{C},
  \quad
  \mathsf{L}_\mathcal{E} = -\mathsf{D}^\dagger \mathsf{D} - \mathsf{C} \mathsf{C}^\dagger,
\end{equation}
and $[\mathsf{L}_\mathcal{C} \mathsf{c}]$, $[\mathsf{L}_\mathcal{V} \mathsf{v}]$, $[\mathsf{L}_\mathcal{F} \mathsf{f}]^{(\ell)}$, and $[\mathsf{L}_\mathcal{E} \mathsf{e}]^{(\ell)}$ can be computed as
\begin{equation}
  (\Delta x)^2 [ \mathsf{L} \mathsf{a} ]_{i,j,k}
    = - 6 \mathsf{a}_{i,j,k}
    + \sum_{q\in\{-1,1\}} \left(
      \mathsf{a}_{i+q,j,k} + \mathsf{a}_{i,j+q,k} + \mathsf{a}_{i,j,k+q} \right).
\end{equation}

\item\emph{Discrete nonlinear operator}: $\tilde{\mathsf{N}} : \mathbb{R}^\mathcal{F} \mapsto \mathbb{R}^\mathcal{F}$, where %
\begin{equation}
  [\tilde{\mathsf{N}}(\mathsf{f})]_{i,j,k}^{(1)} =
    \frac{1}{4} \sum_{q\in\{-1,0\}} \left[
      \mathsf{e}_{i,j,k+q}^{(2)} \left(
        \mathsf{f}_{i,j,k+q}^{(3)} + \mathsf{f}_{i+1,j,k+q}^{(3)} \right)
      - \mathsf{e}_{i,j+q,k}^{(3)} \left(
        \mathsf{f}_{i,j+q,k}^{(2)} + \mathsf{f}_{i+1,j+q,k}^{(2)} \right) \right],
  \label{eq:opr_def_nonlin}
\end{equation}
and $\mathsf{e} = \mathsf{C} \mathsf{f}$.%
\footnote{%
The discrete nonlinear operator presented here is based on the discretization of the convective term in its rotational form, i.e. $\boldsymbol{\omega}\times\mathbf{u} - \frac{1}{2} \nabla (\mathbf{u}\cdot\mathbf{u})$, following the technique described in \citet{zhang2002}.
As discussed in Section~\ref{sec:spatial_discrete}, $\tilde{\mathsf{N}}(\mathsf{f})$ is an approximation of $(\nabla \times \mathbf{f}) \times \mathbf{f}$.
}

\item\emph{Linearized discrete nonlinear operator}:
the linearized form of $\tilde{\mathsf{N}}(\mathsf{f})$ about a constant uniform base flow, $\mathsf{f}_\text{base}(\mathbf{n},t)=\mathbf{f}_\text{base}$, is given by $\mathsf{M}\mathsf{f}^\prime = [\mathsf{K}(\mathbf{f}_\text{base})] \mathsf{C} \mathsf{f}^\prime$, where $\mathsf{f}^\prime = \mathsf{f} - \mathsf{f}_\text{base}$ and
\begin{equation}
    \mathsf{K}(\mathbf{f}_\text{base}) : \mathbb{R}^{\mathcal{E}} \mapsto \mathbb{R}^{\mathcal{F}}, \,
    [ [ \mathsf{K}(\mathbf{f}_\text{base}) ] \mathsf{e}^\prime ]_{i,j,k}^{(1)}
      = \frac{1}{2} \sum_{q\in\{-1,0\}} \left(
      f_\text{base}^{(3)} \mathsf{e}_{i,j,k+q}^{(2)}
      - f_\text{base}^{(2)} \mathsf{e}_{i,j+q,k}^{(3)} \right).
    \label{eq:lin_nonlin}
\end{equation}

\end{itemize}

Discussions regarding the properties of discrete operators are often facilitated by using a block vector/matrix notation to describe the grid functions and linear operators.
Consider the grid spaces $\mathcal{X}$ and $\mathcal{Y}$ corresponding to either $\mathcal{F}$ or $\mathcal{E}$.
Using block vector notation, a vector-valued grid function $\mathsf{x}\in\mathbb{R}^\mathcal{X}$ is expressed as
\begin{equation}
  \mathsf{x} = \mathbb{S}_\mathcal{X}
    [ \bar{\mathsf{x}}_1, \bar{\mathsf{x}}_2, \bar{\mathsf{x}}_3 ]^\dagger,
\end{equation}
where the $q$-th block, $\bar{\mathsf{x}}_q$, corresponds to the values of the $q$-th component of $\mathsf{x}$.
Each $\mathsf{x}_q$ is a scalar real-valued grid function defined on an infinite Cartesian reference grid, which we denote by $\mathbb{R}^\Lambda$.%
\footnote{%
Grid functions in $\mathbb{R}^\Lambda$ can also be regarded as functions mapping $\mathbb{Z}^{3}$ to $\mathsf{R}$.
}
The shift operator $\mathbb{S}_\mathcal{X}:\mathbb{R}^\Lambda\mapsto\mathbb{R}^\mathcal{X}$ is used to transfer, or ``shift'', the values of grid functions defined on $\mathbb{R}^\Lambda$ to $\mathbb{R}^\mathcal{X}$ such that $[\mathsf{x}]^{(q)}(\mathbf{n}) = \bar{\mathsf{x}}_q(\mathbf{n})$.
Similarly, the transpose of $\mathbb{S}_\mathcal{X}$, denoted by $\mathbb{S}_\mathcal{X}^\dagger$, transfers values of grid functions defined on $\mathbb{R}^\mathcal{X}$ to $\mathbb{R}^\Lambda$.
The block vector notation and shift operators readily extend to the case of linear operators.
Using block matrix notation, a discrete linear operator $\mathsf{T}:\mathbb{R}^\mathcal{X}\mapsto\mathbb{R}^\mathcal{Y}$ is expressed as
\begin{equation}
  \mathsf{T} = \mathbb{S}_\mathcal{Y} [ \overline{\mathsf{T}}_{i,j} ] \mathbb{S}_\mathcal{X}^\dagger,\quad i,j = 1, 2, 3,
\end{equation}
where $\overline{\mathsf{T}}_{i,j} : \mathbb{R}^{\Lambda} \mapsto \mathbb{R}^{\Lambda}$.

We now turn our attention to the Fourier representations of grid functions and discrete linear operators.
Consider the Fourier series, $\mathfrak{F}$, and the inverse Fourier transform, $\mathfrak{F}^{-1}$, given by:
\begin{equation}
  [ \mathfrak{F} \bar{\mathsf{u}} ](\boldsymbol{\xi})
    = \sum_{\mathbf{m}\in\mathbb{Z}^3} e^{i \mathbf{m} \cdot \boldsymbol{\xi}}
      \bar{\mathsf{u}}, \quad
  [ \mathfrak{F}^{-1} \hat{\mathsf{u}} ](\mathbf{m})
    = \frac{1}{(2\pi)^3} \int_{\boldsymbol{\xi}\in\Pi}
      e^{-i \boldsymbol{\xi} \cdot \mathbf{m}} \bar{\mathsf{u}}( \boldsymbol{\xi} )\,
      d\boldsymbol{\xi},
\end{equation}
respectively, where $\Pi = (-\pi,\pi)^3$, $\mathsf{u}:\mathbb{Z}^3\mapsto\mathbb{R}$, and $\hat{\mathsf{u}}:\Pi\mapsto\mathbb{C}$.
Using block matrix notation, we extend $\mathfrak{F}$ and $\mathfrak{F}^{-1}$ to the case of grid functions in $\mathbb{R}^{X}$ by defining:
\begin{equation}
  \mathfrak{F}_\mathcal{X} =
    \text{diag}(\mathfrak{F},\mathfrak{F},\mathfrak{F}) \mathbb{S}_\mathcal{X}, \quad
  \mathfrak{F}^{-1}_\mathcal{X} =
    \mathbb{S}^\dagger_\mathcal{X} \text{diag}(\mathfrak{F},\mathfrak{F},\mathfrak{F}).
  \label{eq:grid_fourier}
\end{equation}

Next, let $\Xi$ denote the set of all linear operators $\overline{\mathsf{Q}} : \mathbb{R}^{\Lambda} \mapsto \mathbb{R}^{\Lambda}$ such that the action of $\overline{\mathsf{Q}}$ on an arbitrary grid function $\bar{\mathsf{u}}\in\mathbb{R}^\Lambda$ is given by
\begin{equation}
  [ \overline{\mathsf{Q}} \bar{\mathsf{u}} ] (\mathbf{n})
    = [ \overline{\mathsf{K}}_\mathsf{Q} * \bar{\mathsf{u}} ] (\mathsf{n})
    = \sum_{\mathbf{m}\in\mathbb{Z}^3} \overline{\mathsf{K}}_\mathsf{Q}(\mathbf{m}-\mathbf{n})
      \bar{\mathsf{u}}(\mathbf{m}),
  \label{eq:kernel_q}
\end{equation}
where $\overline{\mathsf{K}}_\mathsf{Q} : \mathbb{Z}^3 \mapsto \mathbb{R}$ is a well-behaved discrete kernel function.
Any operator belonging to $\Xi$ is diagonalized using $\mathfrak{F}$ and $\mathfrak{F}^{-1}$,
\begin{equation}
  [ \overline{\mathsf{Q}} \bar{\mathsf{u}} ] (\mathbf{n})
    = [ \overline{\mathsf{K}}_\mathsf{Q} * \bar{\mathsf{u}} ] (\mathsf{n})
    = [ \mathfrak{F}^{-1} (\hat{\mathsf{K}}_\mathsf{Q} \hat{\mathsf{u}}) ] (\mathsf{n}),
\end{equation}
where $\hat{\mathsf{K}}_\mathsf{Q} = \mathfrak{F} \mathsf{K}_\mathsf{Q}$ and $\hat{\mathsf{u}} = \mathfrak{F} \mathsf{u}$.
The block operators of all linear operators used in the present method belong to $\Xi$.

\section{Lattice Green's functions representations}
\label{app:lgfs}

The NS-LGF method uses the LGFs $\mathsf{G}_{\mathsf{L}}$ and $\mathsf{G}_{\mathsf{E}}(\alpha)$ to computed the action of $\mathsf{L}^{-1}_\mathcal{Q}$ and $\mathsf{E}_\mathcal{Q}(\alpha)$, respectively.
Fourier and Bessel integrals for $\mathsf{G}_{\mathsf{L}}$ and $\mathsf{G}_{\mathsf{E}}$ are given by
\begin{subequations}
  \begin{align}
    (\Delta x)^2 \mathsf{G}_{\mathsf{L}}(\mathbf{n})
      &= \frac{1}{8\pi^3}
        \int_{\Pi} \frac{ \exp\left( -i \mathbf{n}\cdot\boldsymbol{\xi} \right) }
          { \sigma(\boldsymbol{\xi}) }\,d\boldsymbol{\xi}
      = - \int_{0}^{\infty} e^{-6t} I_{n_1}(2t) I_{n_2}(2t) I_{n_3}(2t)\,dt \\
    [\mathsf{G}_{\mathsf{E}}(\alpha)](\mathbf{n})
      &= \frac{1}{8\pi^3}
        \int_{\Pi} \exp\left( -i \mathbf{n}\cdot\boldsymbol{\xi}
          - \sigma(\boldsymbol{\xi}) \right) \,d\boldsymbol{\xi}
      = e^{-6\alpha} I_{n_1}(2\alpha) I_{n_2}(2\alpha) I_{n_3}(2\alpha)
      \label{eq:lgf_intfac}
  \end{align}
\end{subequations}
where $\sigma(\boldsymbol{\xi}) = 2\cos(\xi_1) + 2\cos(\xi_2) + 2\cos(\xi_3) - 6$, $\Pi=(-\pi,\pi)^3$, and $I_n(z)$ is the modified Bessel function of the first kind of order $n$.

Insights into the approximate behavior of $\mathsf{G}_{\mathsf{L}}(\mathbf{n})$ can be obtained by considering the case of $|\mathbf{n}|\rightarrow\infty$.
Asymptotic expansions in terms of unique rational functions for $\mathsf{G}_{\mathsf{L}}(\mathbf{n})$ are provided in \cite{martinsson2002}.
For example,
\begin{equation}
  (\Delta x)^2 \mathsf{G}_{\mathsf{L}}(\mathbf{n}) =
    - \frac{ 1 } { 4 \pi |\mathbf{n}| }
    - \frac{ n_1^4 + n_2^4 + n_3^4
      - 3 n_1^2 n_2^2 - 3 n_1^2 n_3^2 - 3 n_2^2 n_3^2 }
      { 16 \pi |\mathbf{n}|^7 } + \mathcal{O}\left(|\mathbf{n}|^{-5}\right),
\end{equation}
as $|\mathbf{n}|\rightarrow\infty$.
As expected, the leading order term corresponds to the fundamental solution of the Laplace operator.

Numerical procedures for efficiently evaluating $\mathsf{G}_{\mathsf{L}}(\mathbf{n})$ are provided in \cite{liska2014}.
Values for $[\mathsf{G}_{\mathsf{E}}(\alpha)](\mathbf{n})$ can be readily computed using its Bessel form given by Eq.~\eqref{eq:lgf_intfac}.
Although computing values of $\mathsf{G}_{\mathsf{L}}(\mathbf{n})$ and $[\mathsf{G}_{\mathsf{E}}(\alpha)](\mathbf{n})$ can potentially require a non-trivial number of operations, the LGF-FMM, used to compute the action of $\mathsf{L}^{-1}_\mathcal{Q}$ and $\mathsf{E}_\mathcal{Q}(\alpha)$, employs pre-processing techniques that limit the evaluation of point-wise values of LGFs to once per simulation.

\section{Stability analysis}
\label{app:stability}

Consider the linearization of Eq.~\eqref{eq:dnstp_trans} with respect to $\mathsf{v}$ about a uniform, constant base flow, $\mathsf{v}_\text{base}(\mathbf{n},t)=\tilde{\mathbf{u}}$, for the case of $\mathsf{u}_\infty=0$,
\begin{equation}
  \frac{ d \mathsf{v}^\prime }{ d t }
    = [\mathsf{K}(\tilde{\mathbf{u}})] \mathsf{C} \mathsf{v}^\prime
      + \mathsf{G} \mathsf{b}^\prime,\quad
  \mathsf{G}^\dagger \mathsf{v}^\prime = 0,
  \label{eq:lin_eq}
\end{equation}
where $\mathsf{v} = \mathsf{v}_\text{base} + \mathsf{v}^\prime$ and $\mathsf{K}(\tilde{\mathbf{u}})$ is defined by Eq.~\eqref{eq:lin_nonlin}.%
\footnote{%
It is not necessary to linearize the integrating factors present in Eq.~\eqref{eq:dnstp_trans}, since they can be commuted and made to cancel out after the linearization of $\tilde{\mathsf{N}}$.
}
The stability analysis of Eq.~\eqref{eq:lin_eq} is facilitated by using a null-space approach to transform the original DAE index 2 system to an equivalent ODE,
\begin{equation}
  \frac{ d \mathsf{q} }{ d t }
    = \mathsf{C}^\dagger [\mathsf{K}(\mathsf{v}_\text{base})] \mathsf{q}
  \label{eq:lin_eq_trans}
\end{equation}
where $\mathsf{q} = \mathsf{C} \mathsf{v}^\prime$, $\mathsf{v}^\prime = \mathsf{C}^\dagger \mathsf{s}$, and $\mathsf{L}_\mathcal{E} \mathsf{s} = \mathsf{q}$ with $\mathsf{s}\rightarrow 0$ as $|\mathbf{n}|\rightarrow 0$.
The details regarding the feasibility and equivalence of this transformation will be discussed in Section~\ref{sec:truncation_active}.
It is readily verified that the discrete equations corresponding to the HERK method for Eq.~\eqref{eq:lin_eq} and for Eq.~\eqref{eq:lin_eq_trans} are also equivalent; hence, Eq.~\eqref{eq:lin_eq} and Eq.~\eqref{eq:lin_eq_trans} have the same stability region.

The ODE given by Eq.~\eqref{eq:lin_eq_trans} is diagonalized by the component-wise Fourier series $\mathfrak{F}_\mathcal{E}$, defined by Eq.~\eqref{eq:grid_fourier},
\begin{subequations}
\begin{equation}
  \frac{ d \hat{q}_k }{ d t }
    = \frac{|\tilde{\mathbf{u}}|\Delta t}{\Delta x}
      \sigma( \boldsymbol{\xi} ) \hat{q}_k \,\,\,
    \forall i = 1,2,3,
  \label{eq:lin_eq_trans_diag}
\end{equation}
\begin{equation}
  \sigma( \boldsymbol{\xi} ) = - i \sum_{j=1}^3 \frac{ \tilde{u}_j }
    { |\tilde{\mathbf{u}}| } \sin \xi_i,
  \label{eq:lin_eq_eig}
\end{equation}
\label{eq:lin_eq_trans_diag_full}%
\end{subequations}
where $\boldsymbol{\xi}\in\Pi=(-\pi,\pi)^3$.%
\footnote{%
In order to simplify the expression for $\sigma(\boldsymbol{\xi})$ to the form given by Eq.~\eqref{eq:lin_eq_trans_diag} it is necessary to account for $\mathsf{D} \mathsf{q} = 0$.
}
It follows from Eq.~\eqref{eq:lin_eq_eig} that $\Re(\sigma( \boldsymbol{\xi} ))=0$ and $|\Im(\sigma( \boldsymbol{\xi} ))| \le \sqrt{3}$ for all $\boldsymbol{\xi}\in\Pi$.
As a result, the linear stability Eq.~\eqref{eq:dnstp_trans} is determined by the stability of the scalar ODEs:
\begin{equation}
  \frac{dy}{dt} = i \mu y \quad \forall\mu\in (-\gamma,\gamma),\quad
  \gamma=\sqrt{3}\frac{|\tilde{\mathbf{u}}|\Delta t}{\Delta x}.
  \label{eq:scalar_ode}
\end{equation}

Consider integrating the ODE given by Eq.~\eqref{eq:scalar_ode} using the HERK method.
In the absence of algebraic constraints, an HERK scheme reduces to a standard ERK scheme with the same tableau.
Consequently, the region of absolute stability for the ODE of Eq.~\eqref{eq:scalar_ode} is given by
\begin{equation}
  \Omega = \left\{ \mu \in \mathbb{R} : |R(i\mu)| < 1 \right\},
  \quad
  R(z) = 1 + z \mathbf{b}^\dagger \left(\mathbf{I}-z\mathbf{A}\right)^{-1}\mathbf{e},
  \label{eq:stability_region}
\end{equation}
where $\mathbf{b}$ and $\mathbf{A}$ are defined by Eq.~\eqref{eq:erk_tableau}, and $\mathbf{e}=[\,1,\,1,\dots,\,1\,]$ \cite{hairer1996}.
Eq.~\eqref{eq:stability_region} implies that the IF-HERK method is linearly stable if the following CFL condition is satisfied:
\begin{equation}
  \text{CFL} = \frac{|\tilde{\mathbf{u}}| \Delta t}{\Delta x} < \text{CFL}_{\text{max}},
  \quad
  \text{CFL}_\text{max}=\frac{\mu^*}{\gamma}
  \label{eq:cfl_cond}
\end{equation}
where $\mu^* = \text{sup}\left(\Omega\right)$ depends on the RK coefficients of the scheme.
For all the IF-HERK schemes defined in Eq.~\eqref{eq:ifherk_schemes}, the value of $\text{CFL}_\text{max}$ is unity.

\section{Error estimates for integrating factors operating on truncated source fields}
\label{app:iferror}

In this appendix we provide estimates for the difference between $\mathsf{E}_\mathcal{Q}(\alpha)$ and $\mathsf{M}^{\gamma}_\mathcal{Q} \mathsf{E}_\mathcal{Q}(\alpha)\mathsf{M}^{\gamma}_\mathcal{Q}$ inside $D_\gamma$, which are pertinent to the discussion of Section~\ref{sec:truncation_refresh}.
Consider the constant uniform scalar field $\mathsf{u}\in\mathbb{R}^\mathcal{Q}$ and the domain $D_\gamma$, where $D_\gamma$ is infinite in the $x$- and $y$-directions and semi-infinite in the $z$-direction.
For this simplified case, it is sufficient to consider the 1D problem of computing
\begin{equation}
  \mathsf{y} = \left[ \mathsf{E}^\prime(\alpha)
    - \mathsf{M}^\prime \mathsf{E}^\prime(\alpha) \mathsf{M}^\prime \right] \mathsf{u}
    = \left[ \mathsf{I} - \mathsf{M}^\prime \mathsf{E}^\prime (\alpha) \mathsf{M}^\prime \right] \mathsf{u},
\end{equation}
where $\mathsf{I}$ is the identity operator,
\begin{equation}
  \mathsf{E}^\prime(\alpha) \mathsf{u} = \mathsf{G}_\mathsf{E}^\prime(\alpha) * \mathsf{u}, \quad
  \mathsf{G}^\prime_\mathsf{E}(n) = e^{-2\alpha} I_n(2\alpha),
\end{equation}
and
\begin{equation}
  [ \mathsf{M}^\prime \mathsf{u} ](k) = \left\{
    \begin{array}{cc}
      \mathsf{u}(k) & \text{if}\,\,k>0\\
      0 & \text{otherwise}
    \end{array} \right. .
\end{equation}
As a result, the magnitude of the normalized difference, $\mathsf{d}$, at $k>0$ is given by
\begin{equation}
  \mathsf{d}(k) = \frac{\mathsf{y}(k)}{|u|} = \sum_{j=0}^{\infty} e^{-2\alpha} I_{k-j+1}(2\alpha),
  \label{eq:if_finite_error}
\end{equation}
where $|u|$ is the magnitude of the uniform field $\mathsf{u}$.
Numerical approximations for $\mathsf{d}(k)$ are obtained by truncating the infinite sum of Eq.~\eqref{eq:if_finite_error} to a finite number of terms, $N$, such that $I_{k-N+1}(2\alpha)/I_{k+1}(2\alpha)$ is less than a prescribed value.%
\footnote{%
For a fixed $z>0$, $I_{n}(z)$ decreases as $n$ increases. For a fixed $z>0$, $I_{n}(z)$ decays faster than any exponential as $n\rightarrow\infty$.
}

As discussed in Section~\ref{sec:truncation_refresh}, the current implementation of the NS-LGF method uses Eq.~\eqref{eq:if_finite_error} to estimate the error associated with approximating $\mathsf{E}_\mathcal{Q}(\alpha)\mathsf{u}$ by $\mathsf{M}^{\text{xsoln}}_\mathcal{Q} \mathsf{E}_\mathcal{Q}(\alpha)\mathsf{M}^{\text{xsoln}}_\mathcal{Q}\mathsf{u}$, where $\mathsf{u}$ is the velocity perturbation field.
For this case, $|u|$ in Eq.~\eqref{eq:if_finite_error} is set to be the maximum value of any component of $\mathsf{u}$ in $D_\text{soln}$.
Numerical experiments of flows similar to those considered in Section~\ref{sec:verif} demonstrate that this technique leads to fairly conservative error estimates; in all experiments the actual error was less than 10\% of the estimated error.
Tighter error bounds that account for the domain shape and the distribution of $\mathsf{u}$ can potentially be obtained, but are not explored in the present work.

\bibliographystyle{model1-num-names}
\bibliography{jcp_nslgf}

\end{document}